\documentclass[aps, prb, twocolumn, amsmath, amssymb, showpacs, nofootinbib,superscriptaddress]{revtex4-2}

\usepackage[utf8]{inputenc}
\usepackage{graphicx,color,xcolor,times,amsmath}
\usepackage[pdftex, colorlinks=true, linkcolor=myblue, citecolor=myblue,urlcolor=myblue]{hyperref}

\DeclareOldFontCommand{\rm}{\normalfont\rmfamily}{\mathrm}

\definecolor{myblue}{rgb}{0,0,1}

\usepackage{soul}\setstcolor{red} 
\newcommand{\iogs}{Laboratoire Charles Fabry, UMR 8501, Institut d'Optique, CNRS, Université Paris-Saclay,\\ 2 Avenue Augustin Fresnel, 91127 Palaiseau Cedex, France}
\newcommand{\ulyon}{Institut Lumière Matière, Universit\'e Claude Bernard Lyon 1, CNRS, Universit\'e de Lyon, 69622 Villeurbanne, France}
\begin{document}
%
\title{Role of Nottingham effect in the heat transfer in extreme near-field regime}
\author{Mauricio~G\'omez~Viloria}
\email{mauricio.gomez-viloria@institutoptique.fr}
\affiliation{\iogs}
\author{Yangyu~Guo}
\affiliation{\ulyon}
\author{Samy~Merabia}
\affiliation{\ulyon}
\author{Philippe~Ben-Abdallah}
\affiliation{\iogs}
\author{Riccardo~Messina}
\email{riccardo.messina@institutoptique.fr}
\affiliation{\iogs}
%
\begin{abstract}
We analyze the heat transfer between two metals separated by a vacuum gap in the extreme near-field regime. In this cross-over regime between conduction and radiation, heat exchanges are mediated by  photon, phonon and electron tunneling. We quantify the relative contribution of these carriers with respect to both the separation distance between the two bodies and the applied bias voltage. In the presence of a weak bias ($V_{\rm b}<100$~mV), electrons and phonons can contribute equally to the heat transfer near contact, while the contribution of photons becomes negligible. On the other hand, for larger bias voltages, electrons play a dominant role. Moreover, we demonstrate that depending on the magnitude of this bias, electrons can either cool down or heat up the hot body by the Nottingham effect. Our results emphasize some inconsistencies in recent experimental results about heat exchanges in the extreme near-field regime and set a road map for future experiments.
\end{abstract}
\maketitle
\section{Introduction}

Two solids at different temperatures which are separated by a vacuum gap exchange heat by radiation. This exchange of thermal photons is limited, in the far-field regime, by the famous Stefan-Boltzmann’s law which sets the power exchanged between two perfect absorbers (i.e., blackbodies) as the upper limit for the energy that two interacting solids can exchange. In the near-field regime (i.e. at separation distances smaller than the thermal wavelengths of solids), the situation radically changes and the power exchanged between these solids can surpass this blackbody limit thanks to the tunneling of evanescent (i.e. non propagative) photons~\cite{Polder}. In particular, when the materials involved support surface resonant modes, such as surface polaritons or a continuum of hyperbolic modes, the exchanged power can overcome the blackbody limit by several orders of magnitude~\cite{Joulain_rev,Volokitin_rev,Biehs_prl,RMP}. This enhancement of heat exchange predicted by Polder and Van Hove in their seminal work, establishing the foundations of fluctuational electrodynamics (FED), has been verified by numerous experiments~\cite{Hu,Shen,Rousseau,Ottens,Kralik}. This result and the possibility to tune the radiative heat flux at the subwavelength scale have opened new possibilities for the development of innovative technologies for nanoscale thermal management~\cite{Latella}, solid-state cooling~\cite{Fan1,Reddy1}, heating-assisted data storage~\cite{Srituravanich,pba_prl}, IR sensing and spectroscopy~\cite{De Wilde,Jones} and have paved the way to a new generation of energy-conversion devices~\cite{DiMatteo,Narayanaswamy,Laroche,Park,Latella2}.
At closer separation distances, when the objects are separated by atomic distances, further dramatic changes occur. Indeed, in this crossover regime between conduction and radiation also called the extreme near-field regime, some effects and new channels for heat transfer, which are not taken into account by the theory of Polder and Van Hove, appear. More specifically, at the atomic scale, the nonlocal optical response of materials must be taken into account to properly describe the radiative exchanges~\cite{fordweber,poc}. Moreover, acoustic phonons and electrons participate in the transfer through tunneling mechanisms.

Beyond its fundamental interest, the understanding of the extreme near-field regime is of prime importance for the ongoing miniaturization of thermal management and energy technology. However, to date, this physics remains largely elusive and very few experimental works have been reported~\cite{kittel_2017,reddy_17}. Moreover, these works lead to contradictory conclusions. On the one hand the experiment performed in Kittel's group~\cite{kittel_2017} shows a strong deviation with respect to Polder's predictions and seems to demonstrate an extraordinary large heat flux which is four orders of magnitude larger than the values predicted by the conventional theory of fluctuational electrodynamics. On the other hand, the measurements carried out by Reddy's group seem to perfectly reproduce Polder's predictions, even down to the atomic scale. Due to the lack of a general theory to describe the relative role of photons, phonons and electrons in energy exchanges at this scale, this problem remains today in debate. In this paper we introduce a general framework to describe all channels of heat transport in the extreme near-field regime between two metals, with emphasis on the transfer by electron tunneling which plays a major role at the atomic scale. 
We successively study heat carried by elastic vibrations (tunneling of acoustic phonons), photons (near-field radiative heat transfer), and finally free charges (electron tunneling) in the presence of an external bias voltage. This work allows us, on the one hand, to emphasize some limits in the experimental works carried out in Kittel and Reddy's groups. On the other hand, it allows us to quantify the relative contribution of different energy carriers with respect to the separation distance between two solids and the applied bias voltage, thereby setting a potential road map for future experiments.

During the last four years, some attempts to model heat exchanges at the atomic scale have been performed~\cite{Messina_arxiv,Francoeur1,Francoeur2,guo22}. However, in these works, the Nottingham effect~\cite{xu}, that is, the simultaneous heating of both bodies stemming from tunneling electrons in the presence of a bias voltage, has been totally ignored. Here we include this effect in the definition of the heat flux carried by electrons and demonstrate that it significantly modifies both quantitatively and qualitatively, exchanges in the crossover regime between conduction and radiation. Moreover, unlike the previous works, the tunneling probability of electrons is calculated from a rigorous approach based on the transfer-matrix method applied on a Thomas-Fermi description of the electronic potential barrier. This approach allows us to explore the transfer mediated by electrons with an arbitrary bias voltage applied between the two solids in interaction. 

Our paper is structured as follows. In Sec.~\ref{sec:pp}, we address the geometrical configuration of two parallel planes and discuss the formalism used to describe the contribution to heat exchange associated with the three carriers, namely phonons, photons and electrons. We conclude this section by showing their relative contribution as a function of the separation distance and applied potential bias. In Sec.~\ref{sec:tp}, we employ the Derjaguin approximation to discuss the heat transfer in the extreme near field in the tip--plane configuration, and compare our theoretical prediction to the recent experimental results. Finally, in Sec.~\ref{sec:conclusion}, we draw some conclusions and discuss some possible perspectives.

\section{Contribution of the different carriers in plane--plane configuration}
\label{sec:pp}

\begin{figure}[t]
\includegraphics[width=\linewidth]{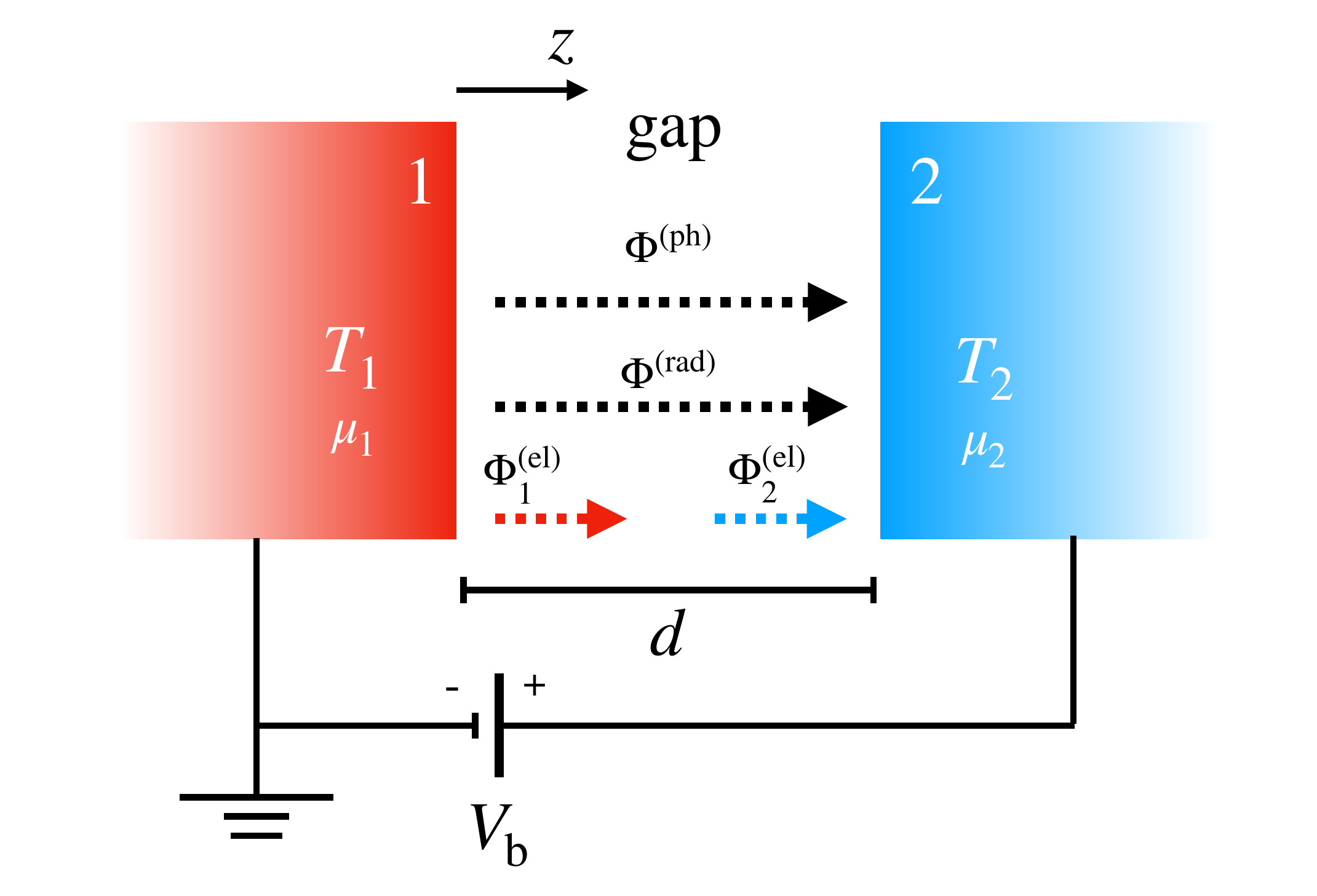}
\caption{\label{fig:sketch}%
Sketch of the system of interest. Two semi-infinite metallic slabs separated by a vacuum gap of thickness $d$ are set at temperatures $T_1$ and $T_2$ and have a chemical potential $\mu_1$ and $\mu_2$.
A bias voltage $V_{\rm b}$ is applied between the two bodies. 
Close to contact (i.e. the extreme near-field regime), electrons (el), phonons (ph) and photons (rad) tunnel through the separation gap. The different arrows associated with the electronic flux reflect the possible non-reciprocity of the latter (see text for details).
}
\end{figure}

In this section, we consider two semi-infinite metallic media with parallel planar surfaces at temperatures $T_1$ and $T_2$, respectively, separated by a vacuum gap of thickness $d$ in the extreme near-field regime, namely below 10\,nm and down to the angstrom range as shown in Fig.~\ref{fig:sketch}.
Throughout this paper we identify medium 1 as the hotter body and medium 2 as the colder body, i.e. $T_1\geq T_2$.
We assume that the total heat flux leaving media $i=1,2$ can be written as a sum of the contribution to the heat flux associated with acoustic phonons (ph), electromagnetic radiation (rad) and electronic tunneling currents (el), as
\begin{subequations}
\label{eq:totalflux}
\begin{equation}
\label{eq:totalflux_1}
\Phi_1=\Phi^{\rm (ph)}+\Phi^{\rm (rad)}+\Phi^{\rm (el)}_1\;,
\end{equation}
\begin{equation}
\label{eq:totalflux_2}
\Phi_2=-\left(\Phi^{\rm (ph)}+\Phi^{\rm (rad)}+\Phi^{\rm (el)}_2\right)\;,
\end{equation}
\end{subequations}
where the $\Phi^{\rm (ph)}$, $\Phi^{\rm (rad)}$, $\Phi^{\mathrm{(el)}}_{i}$ are defined in the following subsections in Eqs.~\eqref{eq:flux_ph}, \eqref{eq:flux_rad} and \eqref{eq:flux_el}, respectively, under the form of a Landauer-like expression.
The assumption that the total heat flux can be separated as the sum of the different contributions is expected to be valid when there are no strong coupling mechanisms between carriers (small bias voltages and temperature difference).
Note that our sign convention in Eqs.~\eqref{eq:totalflux_1} and \eqref{eq:totalflux_2} are such that a positive (negative) heat flux on a given body is an outgoing (ingoing) flux, i.e. it tends to decrease (increase) its temperature.
We also account for the presence of a bias voltage $V_{\rm b}$ between the bodies, affecting the contribution of each of the three heat carriers (ph, rad, el), where $V_{\rm b}>0$ is defined by taking body 2 as the positive electrode as shown in Fig.~\ref{fig:sketch}.
Note also that the electronic contribution to the total heat flux is not necessarily reciprocal, i.e $\Phi_{1}\neq-\Phi_{2}$, a lack of symmetry stemming from the difference in statistics between bosons (ph/rad) and electrons in the presence of a bias voltage, as discussed in Sec.~\ref{sec:el}.
In the following we describe in detail the theoretical framework used to calculate the contribution of each carrier.
As a reference, the numerical values for the physical constants used in our calculations are provided in App.~\ref{app:constants}.

\subsection{Acoustic phonons}
\label{sec:ph}

In crystalline media, phonons are one of the key players of heat conduction in the bulk.
However, near the edges of the sample these quasiparticles only modify its surface since lattice vibrations are only defined inside the material.
The existence of such perturbations of material--vacuum interfaces, along with the presence of forces between the two surfaces, leads to the possibility of a phonon to tunnel across a vacuum gap between the two bodies, paving the way to an additional energy-transfer channel.
This possibility was first considered between metallic media in Refs.~\cite{pendry16,pendry17} within a continuum elastic-medium approach, where van der Waals forces play the role of an interaction mechanism between the two bodies.
The heat flux between the metals was reduced to an integral over the phonon energy of the transmission probability, derived by solving the elastic waves equations.
It was later found that an equivalent approach could be obtained by using the fluctuation-dissipation theorem on the surface displacements~\cite{volokitin19,volokitin20}, in the same spirit of FED.
Within the latter approach, the phonon heat flux can be written as
\begin{align}
\label{eq:flux_ph}
	&\Phi^{(\rm ph)}(T_1,T_2,d)=\\
	\nonumber&2\int_0^{\infty}\frac{\mathrm{d}\omega}{2\pi} \hbar \omega\Delta n_{\rm BE}(\omega,T_1,T_2)\int_0^{\omega/c_{\rm min}} \frac{\mathrm{d} k}{2\pi}\; k\; \mathcal{T}^{\rm (ph)} (k,\omega, d),
\end{align}
where
\begin{equation}
	\Delta n_{\rm BE}(\omega,T_1,T_2)=n_{\rm BE}(\omega,T_1)-n_{\rm BE}(\omega,T_2),
\end{equation}
$n_{\rm BE}(\omega,T_i)=1/[\exp(\hbar \omega/k_{\rm B}T_i)-1]$ being the Bose--Einstein distribution, $k$ the parallel component of the wavevector and $\omega$ the angular frequency of each mode.
The Landauer-like expression \eqref{eq:flux_ph} of the heat flux has three ingredients, namely the energy of each mode $\hbar \omega$, the statistics $\Delta n_{\rm BE}(\omega,T_1,T_2)$ and the transmission probability (taking values between 0 and 1) given by 
\begin{align}
\label{eq:transmission_ph}
	&\mathcal{T}^{\rm (ph)}(\omega,k,d)=\\
	\nonumber& \frac{4 b^2\mathrm{Im}M_1(k,\omega)\mathrm{Im}M_2(k,\omega)}{|[1-aM_1(k,\omega)][1-aM_2(k,\omega)]-b^2M_1(k,\omega)M_2(k,\omega)|^2},
\end{align}
where
\begin{equation}\label{eq:mechanicalM}
	M_i(k,\omega)= \frac{\mathrm{i}}{\rho c_{\mathrm{t},i}^2} \left( \frac{\omega}{c_{\mathrm{t},i}} \right)^2 \frac{k_{\mathrm{l},i}}{(k_{\mathrm{l},i}^2-k^2)^2+4k^2 k_{\mathrm{t},i} k_{\mathrm{l},i}}
\end{equation}
is the mechanical susceptibility~\cite{persson} associated with each medium, in which we have defined a longitudinal acoustic wavevector $k_{\mathrm{l},i}=\sqrt{(\omega/c_{\mathrm{l},i})^2-k^2}$ and a transverse acoustic wavevector $k_{\mathrm{t},i}=\sqrt{(\omega/c_{\mathrm{t},i})^2-k^2}$, related to the longitudinal ($c_{\mathrm{l},i}$) and transverse ($c_{\mathrm{t},i}$) speeds of sound in the media.
The mechanical susceptibility $M_i(k,\omega)$ can be derived by solving the elastic wave equations for a given mode ($k,\omega$)-mode in tensor form~\cite{persson} and extracting the normal-normal component to the surface (the only relevant component for plane--plane symmetry).
The integration bounds with respect to $k$ appearing in Eq.~\eqref{eq:flux_ph} are limited by the function $M_i(k,\omega)$, which vanishes above $\omega/c_{\rm min}$, $c_{\rm min}$ being the smallest transverse speed of sound between the two media. In fact, evanescent transverse acoustic waves do not contribute to the heat flux in the elastic model.
Expression~\eqref{eq:transmission_ph} for the transmission probability stems from a fluctuational approach analogous to the one employed in the context of fluctuational electrodynamics (a derivation can be found in App.~\ref{app:FAD}), where the currents are replaced by surface displacements and $M_i(k,\omega)$ plays a similar role to that of the dielectric permittivity in FED. Note that contrary to the expression for the radiative heat flux (see Sec.~\ref{sec:rad}), the transmission probability in the case of phonons~\eqref{eq:transmission_ph} cannot be separated in terms of phonon polarizations, as longitudinal and transverse waves are coupled by the equations of motion.
Supertransverse waves (having a polarization vector parallel to the surface) are not expected to contribute to the heat transfer~\cite{pendry16,pendry17, volokitin19,volokitin20}.\\

For metals, the terms $a$ and $b$ in the transmission probability \eqref{eq:transmission_ph} are functions of $k$ and $d$, related to the van der Waals force, proportional to the Hamaker constant $H$, and to the electrostatic forces through the bias voltage, yielding
\begin{subequations}
\label{eq:a_and_b}
\begin{equation}
	a=\frac{H}{2\pi d^4}+\epsilon_0 \left(\frac{V_{\rm b}}{d}\right)^2 k \coth(kd),
\end{equation}
and
\begin{equation}
	b=\frac{H}{4\pi d^2}k^2 K_2(kd)+ \epsilon_0 \left(\frac{V_{\rm b}}{d}\right)^2 k \operatorname{csch}(kd),
\end{equation}
\end{subequations}
where $K_2(x)$ is the second modified Bessel function of the second kind and $\operatorname{csch}$ represents the hyperbolic cosecant. 
The dependence on $V_{\rm b}$ in Eq.~\eqref{eq:a_and_b} is independent of the van der Waals force and both $a$ and $b$ are symmetric with respect to a change of sign of $V_{\rm b}$.
In the absence of bias voltage ($V_{\rm b}=0$) the transmission probability in Eq.~\eqref{eq:transmission_ph} is very sensitive to the difference between $a$ and $b$.\\

An upper bound to the phononic contribution to the heat flux can be obtained by setting the transmission to 1 in the expression of the flux, obtaining a Stefan-Boltzmann-like law.
This approximation yields $\Phi^{(\rm ph)}_{\rm max}=(c/c_\mathrm{min})^2\sigma_{\rm SB} (T_1^4-T_2^4)$ for bodies of the same material, where $c$ is the speed of light and $\sigma_{\rm SB}=k_{\rm B}^4\pi^2/60\hbar^3 c^2$ is the Stefan--Boltzmann constant. 
This bound corresponds to a maximal conductance of about $100$ GW m${}^{-2}$ K${}^{-1}$ at 300 K for gold semi-infinite media.\\

\begin{figure}[t]
\includegraphics[width=\linewidth]{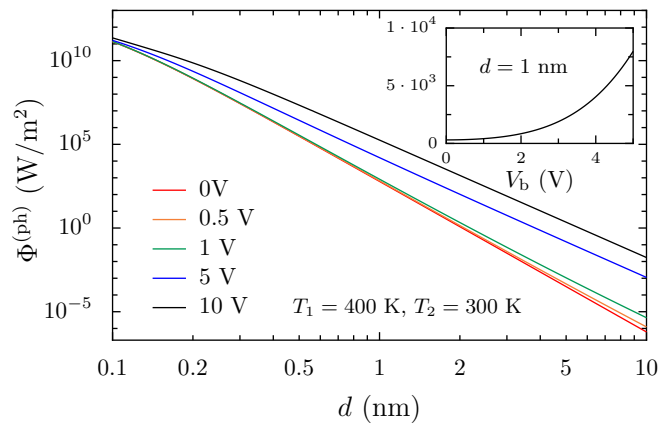}
\caption{\label{fig:pp-ph}%
Phonon heat flux between two semi-infinite gold slabs for various values of the bias voltage $V_{\rm b}$ as a function of the separation distance $d$. Inset: Phonon heat flux between two semi-infinite slabs for $d=1$~nm as a function of $V_{\rm b}$ with a temperature difference of 100 K.%
}
\end{figure}

The phononic heat flow between gold media is shown in Fig.~\ref{fig:pp-ph} as a function of gap distance $d$ and bias $V_{\rm b}$. 
For $d\to 0$, the curves get closer: this results from the fact that the van der Waals interaction becomes dominant and thus the heat flow is less impacted by the bias voltage. 
Above 1\,nm, the dependence of the bias voltage becomes apparent, however the phonon heat flow decreases very rapidly for larger distances.
The slope of the curves for $d>1$ nm is $V_{\rm b}$-dependent.\\

We conclude this section by mentioning that the results obtained within a continuous elastic model for phonons are significantly smaller than the ones recently obtained by means of atomistic calculations, which include the nonequilibrium Green function methods of Refs.~\cite{Francoeur1,Francoeur2} and molecular dynamics in Ref.~\cite{guo22}. A comparison is given in App.~\ref{app:comparison}. However, the limit of validity of the continuous approach to describe the phonon tunneling still remains an open problem which should be analyzed in detail in future works.

\subsection{Photons}
\label{sec:rad}

The theory describing near-field radiative heat transfer was originally treated in the work of Polder and Van Hove~\cite{Polder}.
The theoretical framework they developed, based on FED, predicts a possibly unbounded flux, which can overcome the blackbody limit $\Phi_{\rm BB}=\sigma_{\rm SB}(T_1^4-T_2^4)$ (corresponding to a thermal conductance of 6\,W\,m${}^{-2}$\,K${}^{-1}$ at 300\,K) by several orders of magnitude. More specifically, the first theoretical results predicted a $d^{-2}$ divergence of the heat flux for subnanometer distances.
It was later suggested that such divergence in metals was an artefact resulting from assuming a local dielectric function and could be corrected by taking into account the nonlocal response by accounting for electron-electron interactions~\cite{fordweber,kittel_2005,poc}.\\

The expression for the heat flux carried by electromagnetic waves is analogous to the phonon contribution of Eq.~\eqref{eq:flux_ph}. 
According to FED, it is given by
\begin{align}\label{eq:flux_rad}
	&\Phi^{\rm (rad)}(T_1,T_2,d)=\\
	&\nonumber\int_0^{\infty} \!\frac{\mathrm{d}\omega}{2\pi} \hbar \omega \Delta n_{\rm BE}(\omega,T_1,T_2) \int_0^\infty\!\frac{\mathrm{d} k}{2\pi}\; k\! \sum_{\alpha=\mathrm{s},\mathrm{p}}\mathcal{T}^{\rm (rad)}_\alpha (k,\omega, d),
\end{align}
where the transmission probability,
\begin{equation}\begin{split}
	\label{eq:transmission_rad}
	&\mathcal{T}^{\rm (rad)}_{\alpha}(k,\omega,d)\\
	&\,=\begin{cases} 
		\displaystyle\frac{(1-|r_{\alpha,1}|^2)(1-|r_{\alpha,2}|^2)}{|1-r_{\alpha,1}r_{\alpha,2}\exp(2\mathrm i k_z d )|^2}, & k<\omega/c,\\\vspace{-0.3cm}\\
		\displaystyle\frac{4\,\mathrm{Im}\,r_{\alpha,1}\mathrm{Im}\,r_{\alpha,2}\exp(-2\,\mathrm{Im}\,k_z d)}{|1-r_{\alpha,1}r_{\alpha,2}\exp(-2\,\mathrm{Im}\,k_z d )|^2}, & k\geq \omega/c, \\
	\end{cases}
\end{split}\end{equation}
can now be separated in terms of the two polarizations, given by the transverse electric ($\alpha=\mathrm s$) and transverse magnetic $(\alpha=\mathrm{p})$ contributions, where $k_z=\sqrt{(\omega/c)^2-k^2}$. %
This expression takes into account both propagative ($k<\omega/c$) and evanescent ($k>\omega/c$) waves.\\

Including extreme near-field effects, the reflection coefficients for metals can be written as~\cite{poc,volokitin19}
\begin{subequations} 
\begin{equation}\label{eq:r_s}
		r_{\mathrm s, i} (k,\omega)=\frac{\displaystyle Z_{\mathrm s, i}(k,\omega)-\frac{\omega}{c^2k_z}}{\displaystyle Z_{\mathrm s, i}(k,\omega)+\frac{\omega}{c^2k_z}},
\end{equation}
\begin{equation}
\label{eq:r_p}
	r_{\mathrm p, i}(k,\omega)=\frac{\displaystyle \frac{k_z}{\omega}-Z_{\mathrm p, i}(k,\omega)+\mathrm i\frac{\epsilon_0 k^2}{\omega}\left(\frac{V_{\rm b}}{d}\right)^2M_i(k,\omega)}{\displaystyle\frac{k_z}{\omega}+Z_{\mathrm p, i}(k,\omega)-\mathrm i\frac{\epsilon_0 k^2}{\omega}\left(\frac{V_{\rm b}}{d}\right)^2M_i(k,\omega)},
\end{equation}
\end{subequations}
where $Z_{\alpha, i}$ are the impedances~\cite{fordweber,poc} given by
\begin{subequations}
\begin{align}
	Z_{\mathrm s, i} (k,\omega)=\frac{2\mathrm i}{\pi \omega}\int_0^\infty \frac{\mathrm{d}q_z}{\epsilon_{\mathrm t, i}(K, \omega)-(cK/\omega)^2},
\end{align}
\begin{align}
		&\nonumber Z_{\mathrm p, i} (k,\omega)\\
		&\,=\frac{2\mathrm i}{\pi \omega}\int_0^\infty\frac{\mathrm{d}q_z}{K}\left[\frac{q_z^2}{\epsilon_{\mathrm t, i}(K, \omega)-(cK/\omega)^2}+\frac{k^2}{\epsilon_{\mathrm l, i}(K, \omega)}\right],
\end{align}
\end{subequations}
where $K^2=k^2+q_z^2$ and $\epsilon_{\mathrm l, i}(K,\omega)$ and $\epsilon_{\mathrm t, i}(K,\omega)$ are explicitly defined in App.~\ref{app:nonlocal}.\\

The mechanical susceptibility appears in the expression \eqref{eq:r_p} for p-polarized waves and not in the expression \eqref{eq:r_s} for s-polarized waves.
This modification comes from the acoustic waves in the presence of a bias voltage, as the oscillation of the displacements of the charged surfaces modulates the radiative response \cite{volokitin19,volokitin20}. 
As in the case of phonons, the dependence on the bias voltage of Eq.~\eqref{eq:r_p} is symmetric with respect to the sign of the bias voltage.
When the dielectric function is nonlocal and anisotropic as in the case of a gas of interacting electrons (metals), one has to distinguish between longitudinal $\epsilon_{\mathrm l, i}(\mathbf{K}, \omega)$ and the transverse $\epsilon_{\mathrm t, i}(\mathbf{K}, \omega)$ responses which are functions of both the three-dimensional wavevector $\mathbf{K}$ and the angular frequency $\omega$ of the modes.
Explicit expressions are obtained from the Lindhard theory for jellium in the time-relaxation approximation and are provided in App.~\ref{app:nonlocal}. 
In this work, we also consider the two common limits of Lindhard expressions, the local case and the static case, presented in subsection~\ref{sec:el}, Eq.~\eqref{eq:epsilon_TF}.
Within the local isotropic case, we use the expression given by the Drude model as
\begin{equation}
\label{eq:drude}
	\epsilon_{\rm loc}(\omega)=\epsilon_\infty-\frac{\omega^2_{\rm pl}}{\omega(\omega+\mathrm{i}\Gamma)},
\end{equation}
where $\omega_{\rm pl}$ is the plasma frequency, $\Gamma$ is the damping coefficient and $\epsilon_\infty$ is a constant.\\

For large gap distances and in the case where there is no bias voltage $V_{\rm b}=0$ and $\epsilon_{\mathrm l,i}(K,\omega),\epsilon_{\mathrm t,i}(K,\omega)\to\epsilon_{\rm loc}(\omega)$, we recover the usual Fresnel reflection coefficients.\\

\begin{figure}[t]
\includegraphics[width=\linewidth]{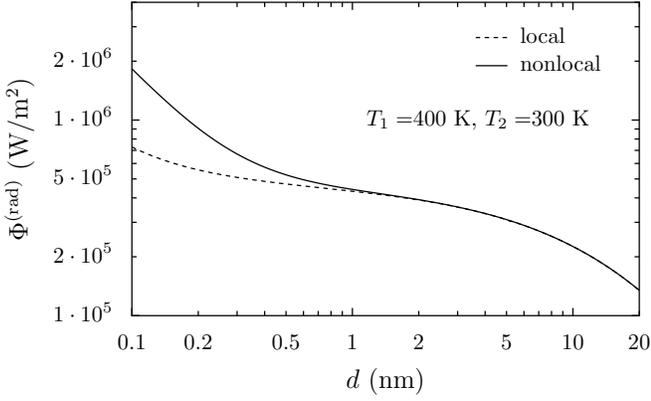}
\caption{\label{fig:pp-rad-nl}%
Local (dashed) and nonlocal (solid) radiative heat flux between two semi-infinite gold slabs as a function of the separation distance $d$.%
}
\end{figure}

The relevance of the nonlocal contribution is shown in Fig.~\ref{fig:pp-rad-nl}, where the nonlocal contribution (dashed line) is compared to the local contribution as a function of the gap distance $d$.
It is known that for metals, the local contribution diverges for small distances.
The nonlocal contribution fixes this issue, as the evanescent contribution from the s-polarized waves saturates close to contact~\cite{poc}.
Nevertheless, there is a region below 1\,nm where the nonlocal contribution exceeds the local contribution, as shown in the figure.
The difficulty to experimentally observe this flux increase is due to the geometry of experiments, where a tip--plane configuration is usually preferred.
We discuss the effects of geometry in Sec.~\ref{sec:tp}.\\

\begin{figure}[t]
\includegraphics[width=\linewidth]{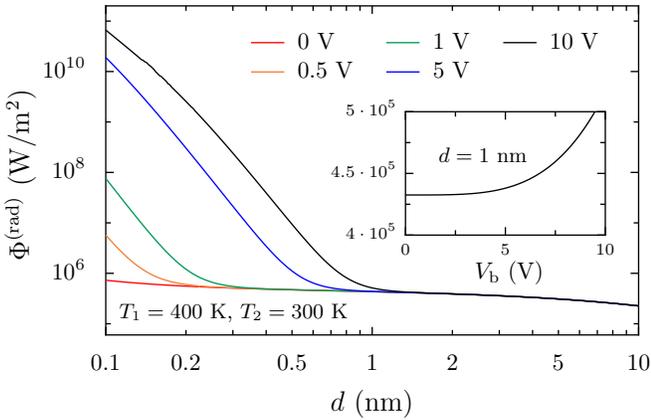}
\caption{\label{fig:pp-rad-bias}%
Radiative heat flux between two semi-infinite gold slabs with respect to the separation distance $d$ for various values of the bias voltage $V_{\rm b}$. Inset: radiative heat flux between two semi-infinite gold slabs with respect to $V_{\rm b}$ for $d=1$~nm.%
}
\end{figure}

The effects of the bias voltage on the radiative contribution were studied in Refs.~\cite{volokitin19,volokitin20}.
While the nonlocal contribution boosted the contribution of the s-polarized waves for a range of gap distances, the bias voltage increases the contribution of p-polarized waves for gaps smaller than 1 nm, as shown in Fig.~\ref{fig:pp-rad-bias}.
For distances above 1\,nm, the standard FED results are recovered.
The increase of the radiative contribution decays slower than the phonon contribution and is also bias dependent.

\subsection{Electrons}
\label{sec:el}

In the experiments described in Refs~\cite{kittel_2017,reddy_17}, electronic currents are measured at separation distances in the nanometer range.
These currents are related to tunneling electrons that can jump from the tip to the sample due to both temperature and voltage biases.
The standard expression for the electric tunneling current density between two semi-infinite media is given by~\cite{simmons,xu}
\begin{align}
	\label{eq:current}
	J=-\frac{e m_{\rm e}}{2\pi^2 \hbar^3}&\int_{\tilde{E}}^{\infty}\mathrm{d}E_z\int_0^{\infty}\mathrm{d}E_{\perp}\\
	\nonumber&\times \Delta n_{\rm FD}(E,T_1,T_2,\mu_1,\mu_2) \mathcal{T}^{\rm (el)}(E_z,V_{\rm b})
\end{align}
where $-e$ is the electron electric charge, $m_{\rm e}$ is the mass of the electron, $E=E_\perp+E_z$ is the total kinetic energy of an electron decomposed in contributions stemming from velocities perpendicular and parallel to the surface, $\tilde{E}=\max(0,-eV_{\rm b})$ and 
\begin{equation}\label{eq:fermidirac}\begin{split}
	\Delta n_{\rm FD}(E,T_1,T_2,\mu_1,\mu_2)&=n_{\rm FD}(E,T_1,\mu_1)\\
	&\quad-n_{\rm FD}(E,T_2,\mu_2),
\end{split}\end{equation}
$n_{\rm FD}(E,T_i,\mu_i)=1/[\exp([E-\mu_i]/k_{\rm B}T_i)+1]$ being the Fermi-Dirac distribution that depends on both temperature $T_i$ and chemical potential $\mu_i$.\\

Close to zero temperature, assuming a degenerate electron gas limit, the current density leaving one body is bounded (in the case of two identical metals) by $J_{\rm max}=e m_{\rm e}E_{\rm F}^2/4\pi^2 \hbar^3$ (about $2.4\times10^{15}$~A/m${}^2$ for gold), where $E_{\rm F}$ is the Fermi energy.\\

The transmission probability $\mathcal{T}^{(\rm el)}(E,V_{\rm b})$ in Eq.~\eqref{eq:current} has only a few analytical solutions for specific electronic barrier shapes.
The semiclassical formula given by the Wentzel--Kramers--Brillouin (WKB) method~\cite{WKB} allows us to estimate the transmission probability through a smoothly varying electronic barrier as
\begin{align}
\label{eq:WKB}
	\mathcal{T}^{\rm (el)}_{\rm WKB}&(E_z,V_{\rm b})=\\
	\nonumber &\exp\left( -\frac{2\sqrt{2m_{\rm e}}}{\hbar}\int_{z_1}^{z_2} \mathrm d z\;\sqrt{U(z, V_{\rm b})-E_z}\right),
\end{align}
where the integration is usually performed between the zeros of the integrand $z_1$ and $z_2$, in the region where the electronic barrier height $U(z,V_{\rm b})$ is larger than the energy $E_z$.
We define the $z$-axis as shown in Fig.~\ref{fig:sketch} and $z=0$ is located at the interface between media 1 and vacuum.
For the temperature of interest, we can safely neglect the temperature dependence of the chemical potential and identify $\mu_1=E_{\rm F}$.
In the case of two metallic media, electrons need to overcome a work function to leave the metal but also the forces generated by the induced image charges in the surfaces of the two media.
Using the image method, the solution to the (local) Poisson's equation for an electron between two identical perfect metals within a classical approach is given by~\cite{simmons}
\begin{equation}
\label{eq:image_potential}
	U_{\rm cl}(z)=E_{\rm F}+W+\frac{e^2}{16\pi \epsilon_0 d } \left[\Psi(z/d)+\Psi(1-z/d)+2\gamma \right],
\end{equation}
where $W$ is the average work function of the media, $\epsilon_0$ is the vacuum permittivity, $\Psi(z)$ is the digamma function and $\gamma$ is the Euler-Mascheroni constant.
The WKB method in combination with the classical image potential has been used many times to describe tunneling phenomena~\cite{simmons,xu,hishinuma,Messina_arxiv,Francoeur1}. 
Nevertheless this methodology, when applied to a plane--plane geometry, is troublesome in three aspects: (i) the work function has to be determined experimentally or through \textit{ab-initio} calculations, (ii) the classical image potential is divergent for $z=0$ and $z=d$, far from the assumptions of a smoothly varying potential, and (iii) the one-dimensional $1/z$ divergences lead to exactly impenetrable barriers ($\mathcal{T}^{(\rm el)}=0$)~\cite{impenetrable}.
The last two difficulties are usually avoided by translating the image planes inside the metals by a distance of a few angstroms based on density functional theory (DFT) calculations.\\

Bardeen applied the Hartree-Fock method to self-consistently calculate the electron density between two metals, which provided a smooth and continuous electronic barrier~\cite{bardeen}.
In order to screen the effects of external charges, electrons create the so called "double layer" where a fraction of the electrons leaks into the vacuum, while leaving an effective positive charged background near the surface of the metal.\\

The only way to obtain a regularized electronic barrier consists in introducing electron-electron interactions, as in the Lindhard theory of metals that was already invoked for the nonlocal effects of the photonic contribution.
For the case of electrons, Kohn-Sham calculations show that accounting just for exchange interactions leads to a slight overestimation of the electronic barrier in the case of a jellium and one would need to self-consistently add correlation functionals to recover the result of the DFT potential~\cite{eguiluzPRL}.
In this work, we limit ourselves to provide an estimation of the electronic heat flux with a regularized potential using an approximate solution ignoring the effects of the correlation potential.
In the static long-wavelength limit $\omega=0$, we can recover Thomas--Fermi screening theory~\cite{ashcroft}, where Lindhard dielectric functions approximate to
\begin{equation}
\label{eq:epsilon_TF}
	\epsilon_{\rm TF}(K)=1+\frac{ k_{\rm TF}^2}{K^2},
\end{equation}
where $K$ is the magnitude of the wavevector inside the metal, $k_{\rm TF}=\sqrt{e^2 m_{\rm e} k_{\rm F}/\pi^2 \hbar^2\epsilon_0}$ is the inverse of the Thomas--Fermi screening length (about 17 nm${}^{-1}$ for gold), and $k_{\rm F}=\sqrt{2m E_{\rm F}/\hbar^2}$ is the Fermi wavevector (about 12 nm${}^{-1}$ for gold).\\ 

\begin{figure}[t]
\includegraphics[width=\linewidth]{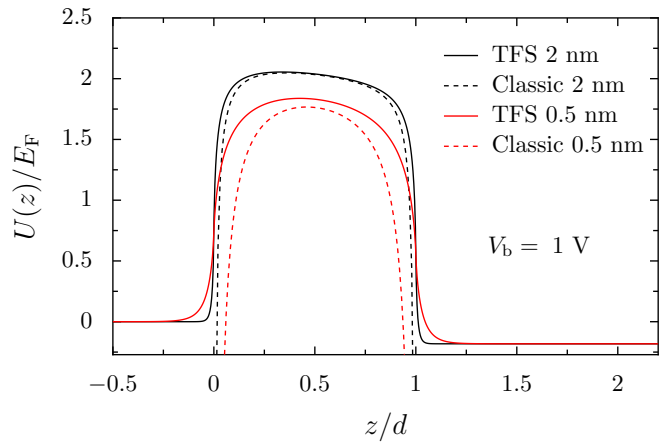}
\caption{\label{fig:potentialbarrier}%
Electronic potential barrier $U(z)$ normalized by the Fermi energy $E_{\rm F}$ between two semi-infinite gold slabs with respect to the position $z$ (in units of the gap thickness $d$) for a bias voltage $V_{\rm b}=1$\,V. The dashed lines represent the classical image potential; the solid lines represent the regularized potential in the Thomas-Fermi approximation. We set the work function of Eq.~\eqref{eq:image_potential} to $W=k_{\rm TF}e^2/8\pi\epsilon_0-E_{\rm F}$ in order to have the same energy reference.
}
\end{figure}

Using a Green's function method, it has been shown that it is possible to solve the nonlocal Poisson's equation in a jellium model in agreement with Thomas--Fermi screening (TFA)~\cite{ilchenko80,ilchenko01}. 
The full electronic potential used in this work is given by
\begin{equation}
\label{eq:regularizedU}
	U(z,V_{\rm b})=\left\{\begin{matrix}
	\displaystyle U_1(z), & z\leq0\\
	\vspace{-0.1cm} & \\
	\displaystyle U_{\rm gap}(z) - \frac zd eV_{\rm b}, & 0<z<d\\
	\vspace{-0.1cm} & \\
	\displaystyle U_2(z)-eV_{\rm b}, &z\geq d
	\end{matrix}\right. ,
\end{equation}
where the expression for $U_s(z)$ ($s=1,2,\mathrm{gap}$) at $V_{\rm b}=0$ is provided in terms of integrals detailed in App.~\ref{app:TFapp}.
In Eq.~\eqref{eq:regularizedU} we have added the effects of the bias voltage linearly inside the gap region, an approximation valid for small $V_{\rm b}<E_{\rm F}$~\cite{ilchenko80}.
The regularized potential is depicted in Fig.~\ref{fig:potentialbarrier} for $V_{\rm b}=1$ V, together with the classical image potential of Eq.~\eqref{eq:image_potential}, for two different separation distances.
Through this method we obtain an electronic barrier of about $2E_{\rm F}$ for gold and large distances (as measured from the bottom of the band) which automatically accounts for the work function.
Below 1 nm the height of the barrier reduces with respect to both the long distance value and the prediction of the classical image potential~\eqref{eq:image_potential}.
The potential reaches different values inside the metal in accordance with the difference in chemical potential due to the bias voltage.
In Fig.~\ref{fig:potentialbarrier} we also show the comparison between two barriers of different gap distances ($d=2$ nm in black and $d=0.5$ nm in red), which shows the effect of reduction of the barrier near contact.
In the experiments this effect of a deviation from the large distances work function was observed in Ref.~\cite{reddy_17}.\\

In order to obtain a more accurate value of the transmission probability, we drop the WKB approximation and calculate $\mathcal{T}^{\rm (el)}(E,V_{\rm b})$ using a transfer matrix method, based on Ref.~\cite{Tmatrixmethod}.
We take slices of the potential and associate a rectangular barrier to each slide with the height of the potential as a function of the coordinate $z$.
The total transmission probability is then calculated as the product of the transmission matrices of all the slices.
This approach has the advantage of taking into account the features of the electronic barrier inside and outside the metal, including the difference in chemical potential due to the bias voltage.\\

\begin{figure}[t]
\includegraphics[width=\linewidth]{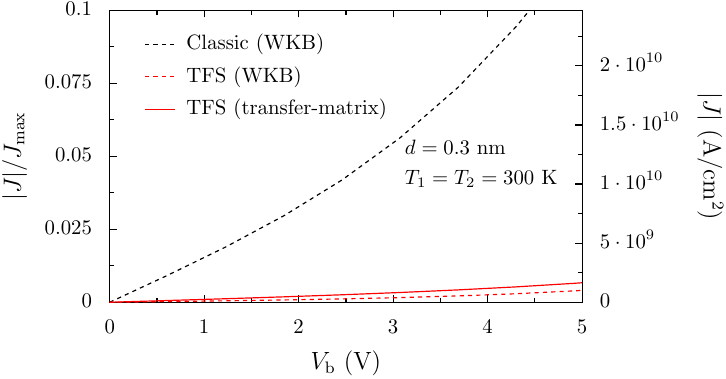}
\caption{\label{fig:pp-el-current}%
Absolute value of the electronic current $J$ (in units of the saturation current $J_{\rm max}$ and in A/cm$^2$) between two semi-infinite gold slabs as a function of the bias voltage $V_{\rm b}$. Dashed lines are calculated using WKB approximation; solid line employs the transfer matrix method. The black line is calculated under the influence of the classic image potential, red lines are calculated for the regularized potential of Eq.~\eqref{eq:regularizedU} within the Thomas--Fermi approximation.}
\end{figure}

\begin{figure}[t]
	\includegraphics[width=\linewidth]{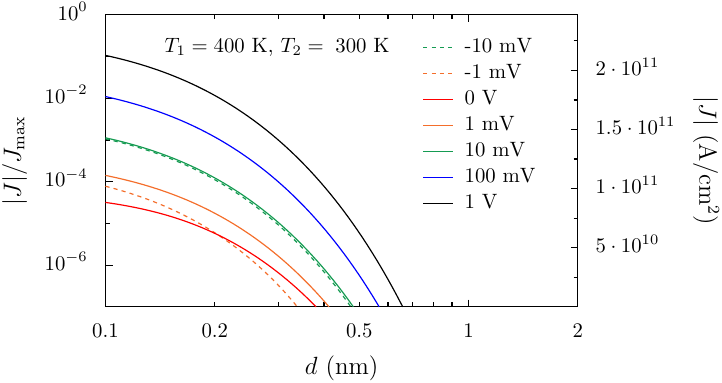}
	\caption{\label{fig:pp-el-current2}%
		Electronic current $J$ (in units of the saturation current $J_{\rm max}$ and in A/cm$^2$) between two semi-infinite gold slabs as a function of the separation distance $d$ and for different bias voltages $V_{\rm b}$. Dashed lines correspond to negative $V_{\rm{b}}$.}
\end{figure}

In Fig. \ref{fig:pp-el-current} we compare the current density between two gold semi-infinite media obtained using the WKB method~\eqref{eq:WKB} and a transfer matrix method for the classical image potential of Eq.~\eqref{eq:image_potential} in the presence of bias voltage and the TFA regularized potential of Eq.~\eqref{eq:regularizedU}.
The dashed lines represent the current using the WKB approximation for the transmission probability of Eq.~\eqref{eq:WKB}, while the solid lines represent the current using the transfer matrix method. 
Due to the difficulties discussed above, the transfer matrix method does not converge when using the classical image potential (not shown).
For small distances the classical potential is significantly smaller in height compared to the TFA potential, and thus the classical potential overestimates the current (black dashed lines) for small distances. 
Comparing the current density using the TFA potential, it can be shown that the WKB calculation (red dashed) underestimates the current compared to the transfer matrix method (red solid).
Contrary to the WKB method that is only sensitive to the shape of the higher part of the barrier, the transfer matrix method has the feature of integrating over the whole shape of the potential, including the difference of potential at the interior of the metals which increases the transmission probability.
The dependence of the current on the separation distance $d$ is shown in Fig.~\ref{fig:pp-el-current2} for different applied bias voltages $V_{\rm b}$. We note, as expected, a strong dependence on the applied bias. Moreover, the current is different from zero even for $V_{\rm b}=0$, for which it is entirely due to the temperature difference. The latter also explains the asymmetry between positive and negative biases, which becomes less and less pronounced when increasing the absolute value of $V_{\rm b}$.

As the number density of electrons in Eq.~\eqref{eq:fermidirac} depends on the chemical potential, we need to define the electronic heat flux carefully.
Contrary to the photons and phonons, for electrons there is a distinction between heat and energy in the thermodynamic sense, and different definitions for the electronic heat flux exist in the literature~\cite{imry82,datta92,buttiker11,Francoeur1}. 
In this work, we write the heat flux leaving media $i=1,2$, following the work of Ref.~\cite{xu}, as
\begin{align}
\label{eq:flux_el}
	\Phi^{(\mathrm{e})}_{i}&(T_1,T_2,d)=\frac{ m_{\rm e}}{2\pi^2 \hbar^3}\int_{\tilde{E}}^{\infty}\mathrm{d}E_z\int_0^{\infty}\mathrm{d}E_{\perp}\\
	\nonumber&\times (E-\mu_i)\Delta n_{\rm FD}(E,T_1,T_2,\mu_1,\mu_2) \mathcal{T}^{\rm (el)}(E_z,V_{\rm b}),
\end{align}
similar to Eqs.~\eqref{eq:flux_ph} and \eqref{eq:flux_rad} for phonons and photons, respectively, where the statistics is now fermionic and the energy in the integrand is replaced by the heat $E-\mu_i$ associated with each reservoir.
The equations for the electronic heat flux are no longer symmetrical and one has to distinguish the heat fluxes for each media. Moreover, the dependence on both temperatures and bias voltage can lead to inversion effects. This can be first shown in the absence of temperature difference ($T_1=T_2$). In this configuration, for $V_b=0$ the energy flux vanishes as expected. The assumption $V_b>0$ corresponds to a shift of the electron energies of body 2 toward lower values. In this scenario, as discussed in Ref.~\cite{xu}, the electrons emitted from body 1 and having energy $E_{\rm F}-eV_{\rm b}<E<E_{\rm F}$ are the ones contributing the most to the energy transfer. When one of these electrons leaves body 1, it is replaced by an electron (provided by the external battery) at energy $E_{\rm F}$, resulting in a net energy flux on body 1 $\Delta E_1 = E_{\rm F} - E$. When the same electron reaches body 2, it provides to it an excess energy $\Delta E_2 = E - E_{\rm F} + e V_{\rm b}$, explaining why both bodies (kept at the same temperature) tend to heat up as a result of the applied potential bias. As soon as a temperature difference $\Delta T = T_1 - T_2 > 0$ is applied, it provides a heating contribution to body 2, which adds to the one already due to the potential bias, and a cooling contribution to body 1, acting thus against the potential bias. This behavior, known as the Nottingham effect~\cite{xu}, manifests itself in a possible inversion of the sign of the heat flux on the hotter body.

Note that Eq.~\eqref{eq:flux_el} ensures that the total power delivered to the system $\Phi^{(\rm el)}_{1}-\Phi^{(\rm el)}_{2}=JV_{\rm b}$ equals the power supplied by the external generator employed to keep the bias voltage constant, for any temperature difference and potential bias.
In thermionic applications, the Nottingham effect has been explored for refrigeration~\cite{hishinuma}. Notice that the electronic heat flux is bounded by $\Phi^{\rm (el)}_{\rm max}=J_{\rm max}E_{\rm F}/3e$.
This value corresponds to about $4\times10^{15}$ W/m${}^2$ for Au-Au tunneling.\\

\begin{figure}[tb]
\includegraphics[width=\linewidth]{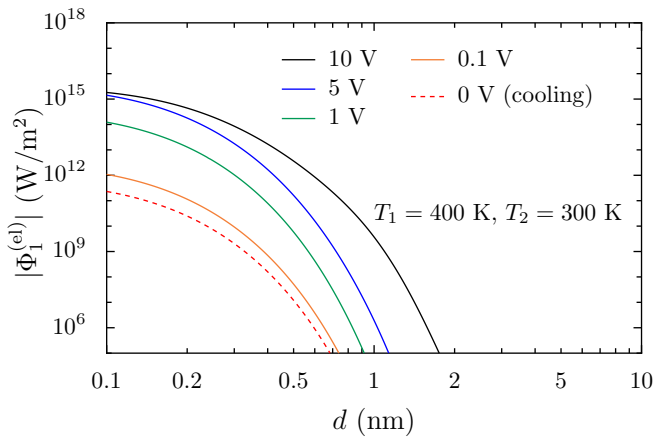}
\caption{\label{fig:pp-el-hflux}%
Absolute value of the electronic heat flux on body 1 in plane--plane configuration for various values of the bias voltage $V_{\rm b}$ as a function of the separation distance $d$. The solid (dashed) lines indicate emitted (receiving) heat from medium 1.%
}
\end{figure}

In Fig.~\ref{fig:pp-el-hflux} we compare the absolute value of the heat flux on the hotter medium 1 for different values of the bias voltage and gap distances.
For large values of $V_{\rm b}$ the flux saturates.
The heat flux for $V_{\rm b}=0$ is in the thermodynamic regime where body 1 tends to cool down by the emission of the electrons, and has the opposite sign of the curves with $V_{\rm b}>0$ (solid lines) which represent heating of body 1.
We stress that our definition \eqref{eq:flux_el} of the electronic heat flux differs from the one employed in Ref.~\cite{Francoeur1}. A comparison of the numerical results obtained within the two approaches is given in App.~\ref{app:comparison}.

\subsection{Comparison}
\begin{figure}[tb]
\includegraphics[width=\linewidth]{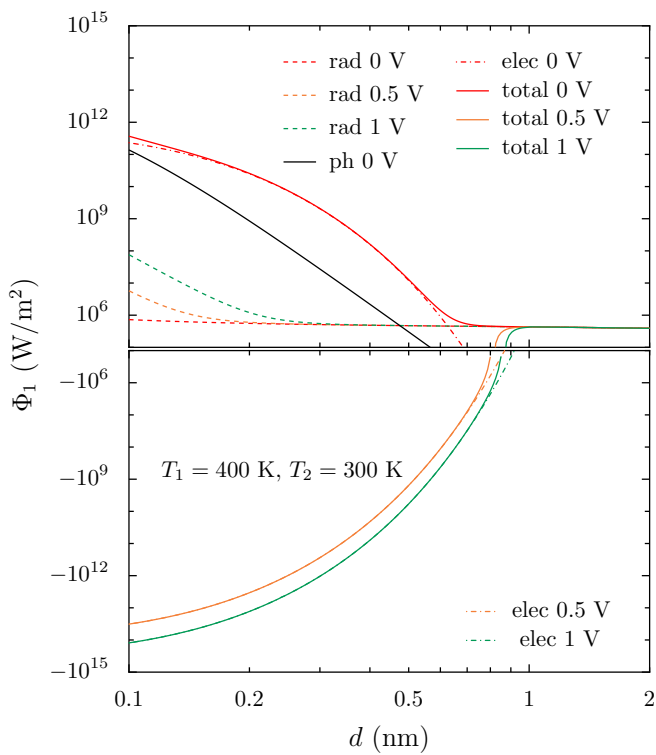}
\caption{\label{fig:pp-all}%
Comparison of the heat flux of the different heat carriers in plane--plane configuration for various values of the bias voltage $V_{\rm b}$ as a function of the separation distance $d$. Positive (negative) values of the heat flux are related to outgoing (incoming) heat on body 1.%
}
\end{figure}

\begin{figure*}[tb]
\includegraphics[width=\linewidth]{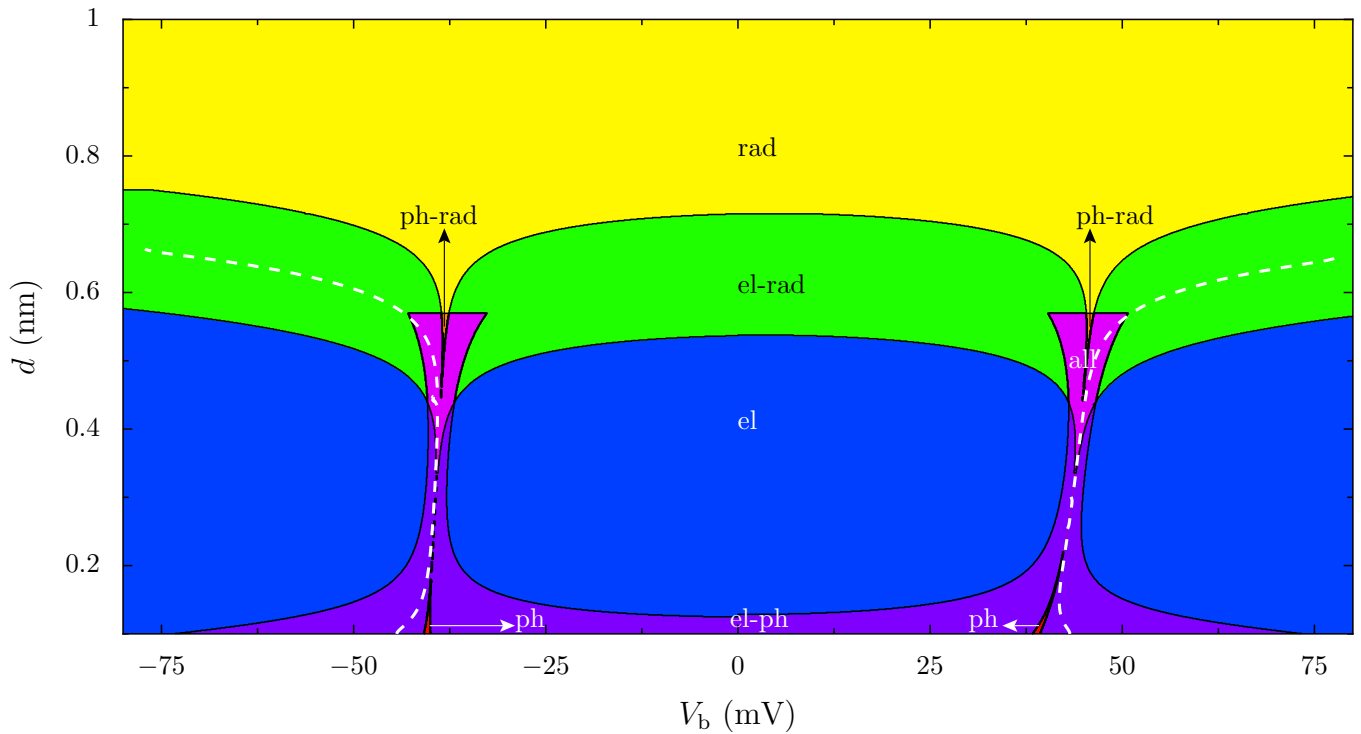}
\caption{\label{fig:colormap}%
Comparison of the absolute value of the heat flux $\Phi_1$ emitted by media 1 in plane--plane configuration for various values of the bias voltage $V_{\rm b}$ and separation distance $d$ for $T_1=400$ K and $T_2=300$ K. The different colors represent the leading heat carriers (ph in red, rad in yellow, el in blue). Regions with two labels (el-rad in green, ph-rad in purple, el-ph in orange) represent regions where the contribution of the two carriers is comparable up to $10\%$ between the two carriers. The label "all" (in pink) represents the regions where all three carriers contribute equally up to a $10\%$ difference. The dashed lines separate the regions where the electronic heat flux has a different sign (heating in the external regions enclosed by the white dashed lines, cooling in the rest of the figure). 
}
\end{figure*}

The comparison of the different contributions to the heat flux (ph, rad, el) is presented in Fig.~\ref{fig:pp-all} with a double logarithmic scale, with positive (negative) values representing the tendency to cool down (heat up) body 1.
The near-field radiative contribution and the phonon contribution are both positive for any separation distance and applied bias voltage, corresponding to heat emitted by body 1.
The comparison between them shows that the phonon contribution dominates over the near-field radiative one below 0.4 nm.
The electronic heat flux is also positive when there is no bias voltage and changes drastically of sign above a few tens of mV.
The solid lines represent the total heat flux as the sum of the three contributions (ph, rad, el). We observe that, while for distances larger than 1\,nm photons are the only relevant carriers, the electronic contribution rapidly dominates below 1\,nm.
These results represent a clear deviation from FED.
The fact that the electronic contribution dominates the heat transfer is no coincidence as the main contribution to the thermal conductivity of metals is electronic in nature (phonons contribute about 10\%).
Near contact (distances around 1\,\AA), the phononic part becomes relevant compared to the electronic contribution as expected.\\
In Fig.~\ref{fig:colormap} we map the dominating contributions to the total heat $\Phi_1$ flowing from the hotter medium 1 as a function of $d$ and $V_{\rm b}$.
Each label in the colored regions of the figure indicates the dominating heat carrier (ph, rad, el).
In some regions, two (el-rad, ph-rad, el-ph) or even the three (all) carriers compete.
The two regions close to the borders of the figure enclosed by the white dashed lines indicate the regions where the heat flux becomes negative (incoming heat).
As expected, at about 0.8 nm radiation clearly dominates and continues to do so for larger distances.
Below 0.5 nm, the region of dominance of the electrons fills the map.
Near the change of sign (dashed white line) and about $\pm40$ mV, the electronic contribution is reduced and the phonons create thin stripes where their contribution competes with the electrons.
Near contact phonons compete with the electrons, as expected, or even become the dominant contribution, but only as long as the bias remains small, i.e. $|V_{\rm b}|<100$ mV. 
The figure is slightly asymmetric with respect to $V_{\rm b} = 0$ due to the temperature difference.
We do not map the heat flowing into media 2 but it is expected to be similar to the intermediate ($-25\,\text{mV}<V_{\rm b}<25\,\text{mV}$) region of Fig.~\ref{fig:colormap} and almost independent of $V_{\rm b}$, without any cooling regions.

The map of carrier contributions to the heat transfer presented in Fig.~\ref{fig:colormap} is one of the main results of this work and could serve as a path to identify experimentally the participation of heat carriers, and in particular of electrons. As a matter of fact, the transition from heating of body 1 ($V_{\rm b} < -50\,\text{mV}$) to cooling ($-50\,\text{mV}<V_{\rm b} < 50\,\text{mV}$) and to heating again ($V_{\rm b} > 50\,\text{mV}$) is a unique signature of the electronic behavior via the Nottingham effect. We also observe that the strong dependence on the separation distance $d$ is such that this feature disappears for larger separation distances ($d>0.7\,\text{nm}$), for which photons tend to dominate the heat transfer. We conclude that an experimental setup able to control the separation distance around 0.5\,nm along with the potential bias $V_{\rm b}$ could allow one to clearly demonstrate the participation of electrons to heat exchange in the extreme near field.

\section{Tip--plane configuration}
\label{sec:tp}

In order to compare our results with the existing experiments, we exploit the Dejarguin or proximity force approximation (PFA), already employed in the context of near-field radiative heat transfer by the authors of Refs.~\cite{kittel_2017,reddy_17}. For a spherical shaped tip of radius $R$, the net power exchanged between the tip and the sample can be written as
\begin{subequations}
\label{eq:PFA}
\begin{equation}
	P_{\rm tip}=(-1)^{i+1}\left(P^{\rm (ph)}+P^{\rm (rad)}+P^{\rm (el)}_{i}\right),
\end{equation}
in agreement with the net flux of Eq.~\eqref{eq:totalflux}, where each term is defined as
\begin{equation}
	P^{\rm (Q)}_i=2\pi\int_0^R \mathrm{d} r \; r\;\Phi^{\rm (Q)}_{i}\!\left(d + R-\sqrt{R^2-r^2}\right),
\end{equation}
\end{subequations}
where $Q\in\{\mathrm{rad,ph,el}\}$ and $i=1,2$ is to be chosen depending on which media corresponds to the tip and the subindex is dropped when discussing phonons and radiation. Analogously, the PFA expression for the electric current reads
\begin{equation}
	I=2\pi\int_0^R \mathrm{d} r \; r\;J\!\left(d + R-\sqrt{R^2-r^2}\right)
\end{equation}
where $J$ is the tunneling current density defined in Eq.\eqref{eq:current}.

\begin{figure}[thb]
\includegraphics[width=\linewidth]{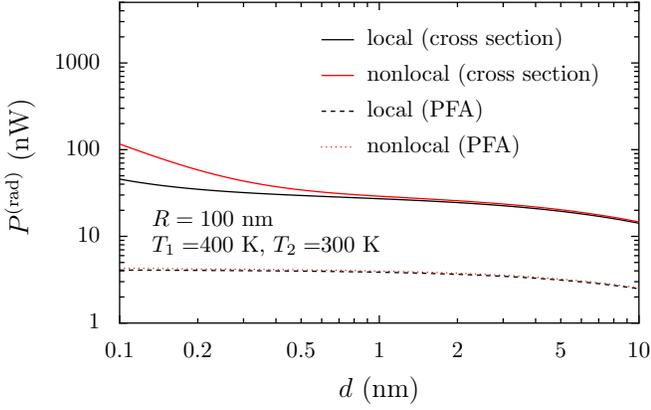}
\caption{\label{fig:pfa-rad-nl}%
Comparison between local (red) and nonlocal (black) radiative power emitted in tip--plane configuration as a function of the separation distance $d$ for a spherical tip of radius $R=100$~nm. The dashed line is calculated using the proximity force approximation; the solid lines represent the heat flux multiplied by the tip cross-section.%
}
\end{figure}

In the supplemental material of Ref.~\cite{kittel_2017} it was already found that the nonlocal effects were not observable in Au-Au experiments with a tip--plane configuration using the PFA.
We provide further evidence of this point in Fig.~\ref{fig:pfa-rad-nl}, where we plot the power emitted by the tip as a function of the gap by multiplying $\Phi^{\rm (rad)}$ by the cross-section of the tip $\pi R^2$ (solid lines), and using the PFA (dashed) lines, in the presence of local and nonlocal effects.
When looking at the flux times the cross-section, we observe that the nonlocal radiative heat flux increases with respect to the local theory in a region below 1 nm, as we have already shown in Fig.~\ref{fig:pp-rad-nl}. 
However, when looking at the PFA curves in Fig.~\ref{fig:pfa-rad-nl}, we observe that the radiative flux is almost constant and it is not possible to distinguish the nonlocal estimate from the local one.
The distance dependence of the radiative flux disappears by integrating over the tip, which translates into a reduction of 1 to 3 orders of magnitude with respect to the solid curves.
The region of interest for the nonlocal effects represent a small area at the edge of the tip and thus gives a negligible contribution to the integral given in Eq.~\eqref{eq:PFA}.\\

\begin{figure}[t]
\includegraphics[width=\linewidth]{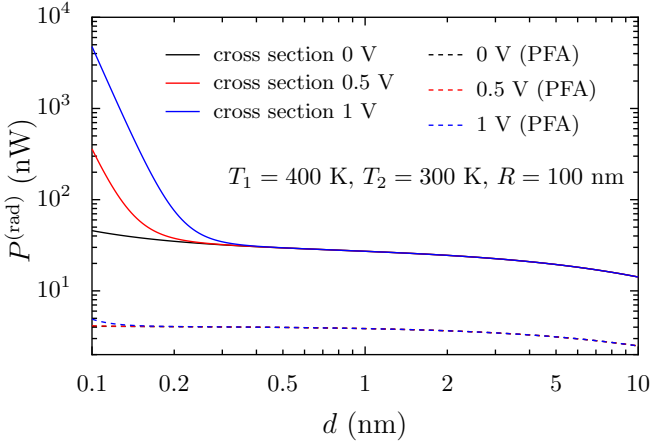}
\caption{\label{fig:pfa-rad-bias}%
Comparison between radiative power exchanged in a tip--plane configuration for different values of the bias voltage $V_{\rm b}$ as a function of the separation distance $d$ for a spherical tip of radius $R=30$~nm. The dashed lines represent the result obtained using the proximity force approximation while the solid lines denote the heat flux multiplied by the tip cross-section.%
}
\end{figure}

While the contribution of the bias voltage to the radiative heat flux can exceed the influence of the nonlocal effects to the radiative transfer, its effects are not striking either.
For this purpose, we also compare the effect of the bias voltage on the transmitted power in flux times cross-section and PFA in Fig.~\ref{fig:pfa-rad-bias}, as a function of gap distance and bias voltage (see Fig.~\ref{fig:pp-rad-bias}).
The quick ascent of the solid curves below 0.5 nm in the presence of the bias with respect to zero bias translates also into almost no difference between the PFA curves (dashed) and the difference between the curves is barely noticeable for the largest bias shown (blue, 1\,V) at vanishing gap distances.
Stronger biases could be used to probe this effect, but aside from being an experimental challenge, the contribution of the other two carriers (ph and el) increases as well and stronger couplings are expected. \\

\begin{figure}[t]
	\includegraphics[width=\linewidth]{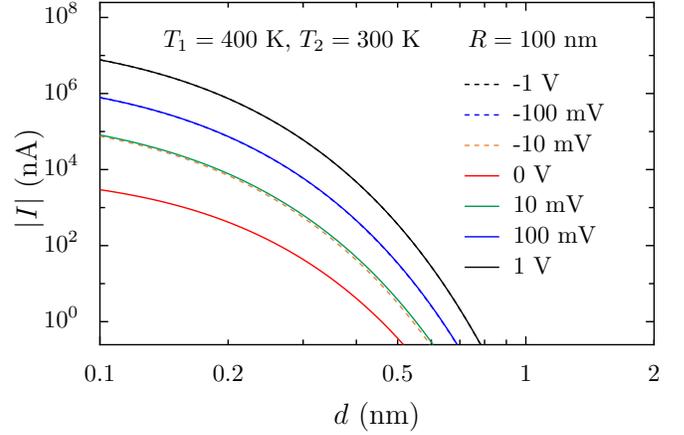}
	\caption{\label{fig:pfa-I}%
		Absolute value of the electric current exchanged as a function of the separation distance in a tip--plane scenario, calculated within the PFA approach, between a gold tip of radius $R=100\,$nm and at temperature $T_1=400\,$K and a gold planar substrate at temperature $T_2=300\,$K. The different curves correspond to different values of the applied bias voltage $V_{\rm b}$. Dashed lines correspond to negative bias voltages and the electronic current has the opposite sign compared to that of solid lines.}
\end{figure}

\begin{figure}[t]
	\includegraphics[width=\linewidth]{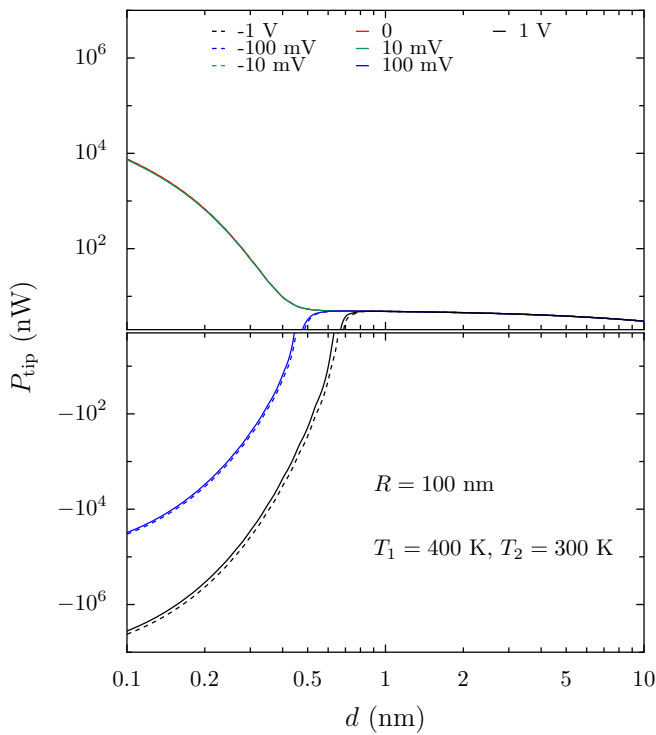}
	\caption{\label{fig:pfa-power}%
		Total power emitted by the tip as a function of the separation distance in a tip--plane scenario, calculated within the PFA approach, between a gold tip of radius $R=100\,$nm and at temperature $T_1=400\,$K and a gold planar substrate at temperature $T_2=300\,$K. The curves at $V_{\rm b}=\pm100\,\text{mV},\pm1\,\text{V}$ switch to take negative values below a separation distance below 1\,nm.}
\end{figure}

In Figs.~\ref{fig:pfa-I} and \ref{fig:pfa-power} we show the current and the total power emitted by the tip calculated using PFA in a tip--plane scenario with a tip radius of $R=100\,$nm. We remark that both quantities are quite sensitive to the applied bias and almost symmetric with respect to a sign change of $V_{\rm b}$. Moreover, note that the tip is heated up for large enough $V_{\rm b}$ (in absolute value) and for small distances.

\begin{figure}[t]
\includegraphics[width=\linewidth]{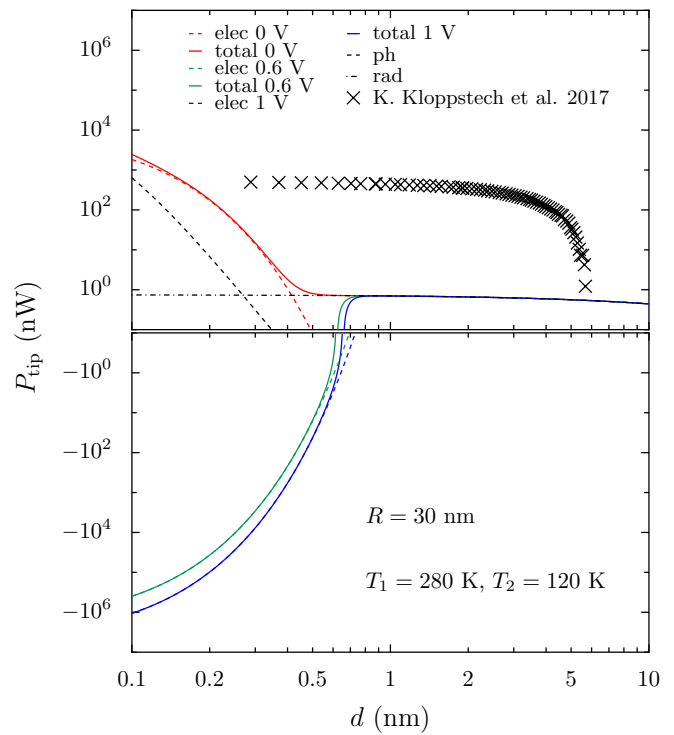}
\caption{\label{fig:pfa-all}%
Comparison of the numerical results with the available experimental data at 600 mV~\cite{kittel_2017}. The curves indicate the power exchanged between the tip and the plane with respect to the separation distance $d$ for various values of the bias voltage $V_{\rm b}$. Calculations are performed in the proximity force approximation with a spherical tip of radius $R=30$~nm.}
\end{figure}

Moving to a theory-experiment comparison, we start by noting that Kloppstech et al.~\cite{kittel_2017} modelled the tip as a sphere with a radius of about 30 nm, while Cui et al.~\cite{reddy_17} had a tip of radius of 150 nm.
Furthermore, the former experiment had a positive temperature bias ($T_1=280$\,K for the tip and $T_2=120$\,K for the sample), while in the latter set-up the bias was inverted ($T_1=445$\,K for the sample and $T_2=315$\,K for the tip).
In Fig.~\ref{fig:pfa-all} we compare our theoretical results stemming from PFA (solid and dashed curves) using Eqs.~\eqref{eq:PFA} to the experimental results of Ref.~\cite{kittel_2017} (symbols $\times$).
While the phononic (black dashed line) and radiative (dash-dotted line) contributions are shown for $V_{\rm b}=0$, the electronic ones are shown for different bias voltages up to 1\,V, including 0.6\,V which is the value applied in the experiment.
The experimental data already diverge from the FED predictions at 7\,nm, much larger than the 1\,nm scale where electrons are expected to contribute.
The authors reported that no electric current was detectable above 1\,nm (below 0.5\,pA), indicating that, even if contamination was present, it was not conductive as it did not enhance the electronic contribution.
However that does not exclude the possibility of an enhancement of the bosonic contributions (ph and rad) or the activation of other heat channels.\\

\begin{figure}[t]
\includegraphics[width=\linewidth]{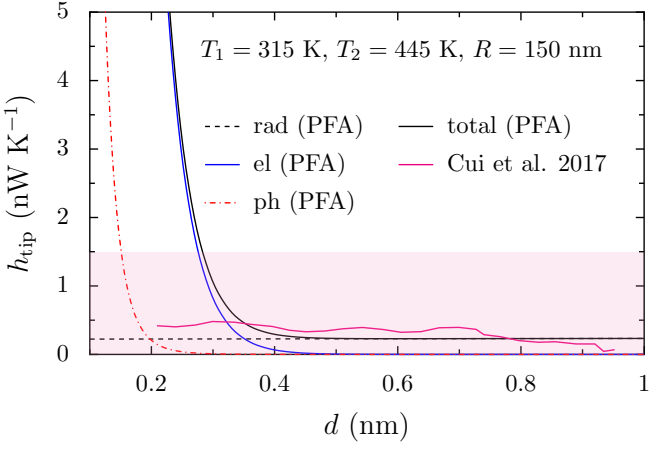}
\caption{\label{fig:pfa-all_2}%
Comparison of the numerical results with the available experimental data~\cite{reddy_17}.
The curves indicate the heat conductance in a tip-plane configuration with respect to the separation distance $d$ with $V_{\rm b}=0$.
Calculations are performed in the proximity force approximation with a spherical tip of radius $R=150$\,nm. Experimental values are indicated by the pink line while the pink background shows the experimental error margin according to Fig. 2d of Ref.~\cite{reddy_17}%
}
\end{figure}

We now focus on the experimental results of Ref.~\cite{reddy_17}, compared to our theoretical predictions in Fig.~\ref{fig:pfa-all_2}.
The figure shows the thermal conductance of the tip $h_{\rm tip}=P_{\rm tip}/\Delta T$ under PFA for $V_{\rm b}=0$, where $\Delta T=T_1-T_2$ is the temperature difference.
The radiative contribution (black dashed line) is in agreement with the experimental data (dark pink solid line) and inside the error range (light pink background), suggesting no deviation from FED.
We also note that the error margin goes to negative values in the experimental data (not shown in the figure).
The behavior of the experimental data suggests an opposite conclusion with respect to Kloppstech et al.~\cite{kittel_2017} where there is a clear deviation from the radiative heat transfer theory.
Our theoretical predictions indicate that the electronic contribution (blue solid line) should overcome the error range for 0.3\,nm and below in Fig.~\ref{fig:pfa-all_2}, which is not detected in the experiment, even if an electronic current is measured at this range.
The phonon contribution (red dash-dotted curve), assuming $V_{\rm b}=0$, starts to be relevant at distances where there are no experimental values.
If a small bias of a few mV was introduced in the experiment, the phonon contribution could be more relevant at larger distances.
A slight rise in the experimental curve (dark pink solid) is seen going to smaller distances but the error margin does not allow one to draw any reliable quantitative conclusion.


\section{Conclusion}
\label{sec:conclusion}

In this work we have shown that, close to the contact, electrons and phonons are the two main contributors to heat exchanges between two metallic solids. At subnanometer distances, the Nottingham effect associated with the electronic current gives rise to a rich thermal behavior (including thermal-rectification effects). Hence, the heating associated to these carriers induces an asymmetry with respect to the sign of the applied bias, as well as in the power received and emitted by the electrodes. When $ |V_{\rm b}|$ is small, the flux mediated by electrons flows in the same direction as the flux mediated both by photons and phonons, cooling down the hotter body. However, when $ |V_{\rm b}|\gg 100$\,mV, the heat flux carried by electrons dominates and is of the opposite sign with respect to the flux carried by the other carriers, simultaneously heating both bodies. This asymmetry, along with the strong behavior as a function of $V_{\rm b}$, could help the experimental identification of the participation of electrons to heat transfer.

The comparison of our theoretical predictions with the existing experimental results show that some problems currently limit their interpretation. In Reddy's experiment~\cite{reddy_17} we have seen that the value of heat flux is below the experimental sensitivity. However, an increase of bias voltage or of temperature difference could probably make the study of heat transfer at the atomic scale possible. On the other hand, in Kittel's experiment~\cite{kittel_2017} the strong spatial shift between the measured flux and the theoretical predictions at distances of the order of $5\,$nm, where neither electrons nor phonons contribute to the transfer, cannot be easily interpreted without considering the presence of molecular contaminations or extra layers such as water layers. Nevertheless, once again, a series of measurement made with various bias voltages could probably help to clarify the current results.

In addition, we think that our work could be useful for future experimental studies of Nottingham effect and to investigate its consequences on the heat exchanges between two metallic solids close to the physical contact.

\begin{acknowledgments}
This research was supported by the French Agence Nationale de la Recherche (ANR), under grant ANR-20-CE05-0021-01 (NearHeat). 
\end{acknowledgments}


\appendix
\section{Reference values for the constants}
\label{app:constants}

In the elastic acoustic wave model, we use $c_{\rm l}=3240$ m/s, $c_{\rm t}=1200$ m/s and $\rho=1.9\times 10^4$ kg/m${}^3$ for gold. 
The van der Waals force between two gold metals is proportional to the Hamaker constant $H=34.76\times 10^{-20}$ J, taken from Ref.~\cite{hamaker}.\\

In Drude's expression for the local dielectric function of Eq.~\eqref{eq:drude} we use $\omega_{\rm pl}=1.36\times10^{16}$ rad/s as the plasma frequency for gold, $\Gamma=1\times10^{14}$\,rad/s and $\epsilon_\infty=9.84$, fitted data~ \cite{Sonnichsen2001} from Ref.~\cite{johnson1972}. 
The Fermi energy is taken as $E_{\rm F}=5.53$ eV for gold.
For the nonlocal calculation, we use $v_{\rm F}=c/214$ from Ref.~\cite{ashcroft}.


\section{Nonlocal dielectric function}
\label{app:nonlocal}
In this appendix we detail the longitudinal and transverse dielectric functions in the random phase approximation (RPA), also known as Lindhard theory for an interacting electron gas, employed in expressions of Sec.~\ref{sec:rad} to describe the the nonlocal radiative heat transfer.\\

For low temperatures compared to the Fermi temperature $T_1,T_2\ll T_{\rm F}$, Lindhard theory together with Mermin's time-relaxation approximation~\cite{fordweber,poc} describes the longitudinal and transverse dielectric functions as 
\begin{subequations}
\begin{equation}
	\epsilon_{\rm l}(K, \omega)=\epsilon_\infty+\frac{3 \omega_{\rm pl}^2}{\omega + \mathrm i \Gamma}\frac{\nu^2 g_{\rm l}(\zeta,\nu)}{\omega+\mathrm i \Gamma \displaystyle \frac{g_{\rm l}(\zeta,\nu)}{g_{\rm l}(\zeta,0)}},
\end{equation}
and
\begin{align}
	\epsilon_{\rm t} (K, \omega)=\epsilon_\infty-\frac{\omega^2_{\rm pl}}{\omega^2(\omega+\mathrm i \Gamma)}&\\
	\nonumber\times\{\omega[g_{\rm t}(\zeta,\nu)-3\zeta^2 g_{\rm l}(\zeta,\nu)]&+\mathrm i \Gamma [g_{\rm t}(\zeta,0)-3\zeta^2 g_{\rm l}(\zeta,0)]\},
\end{align}
\end{subequations}
where $\zeta=K/2k_{\rm F}$ and $\nu=(\omega + \mathrm i \Gamma )/Kv_{\rm F}$, $v_{\rm F}$ the Fermi speed, and 
\begin{subequations}
\begin{align}
	g_{\rm l}(\zeta,\nu)=\frac{1}{2}&+\frac{1-(\zeta-\nu)^2}{8\zeta}\log\left(\frac{\zeta-\nu+1}{\zeta-\nu-1}\right)\\
	\nonumber&+\frac{1-(\zeta+\nu)^2}{8\zeta}\log\left(\frac{\zeta+\nu+1}{\zeta+\nu-1}\right)
\end{align}
and
\begin{align}
	g_{\rm t}(\zeta,\nu)=&\frac{3}{8}(\zeta^2+3\nu^2+1)\\
	\nonumber&-3\frac{[1-(\zeta-\nu)^2]^2}{32\zeta}\log\left(\frac{\zeta-\nu+1}{\zeta-\nu-1}\right)\\
	\nonumber&-3\frac{[1-(z+\nu)^2]^2}{32\zeta}\log\left(\frac{\zeta+\nu+1}{\zeta+\nu-1}\right)
\end{align}
\end{subequations}
are the Lindhard-Mermin functions.
%


\section{Fluctuational acoustodynamics}
\label{app:FAD}

In the fluctuational electrodynamics theory, the components of the currents due to thermal and quantum fluctuations are related to the dielectric function through the fluctuation-dissipation theorem~\cite{volokitin19,volokitin20}. Here we present an acoustic analog for the elastic waves inside each media $i=1,2$.
The $z$-component of the oscillating displacements $u_ i(x,y,t)=u^{\rm fl}_i(x,y,t)+u^{\rm ind}_i(x,y,t)$ at the surface of each media are written as a sum of fluctuating and induced contributions. The induced displacements between the two surfaces $u^{\rm ind}_i$ originate from the van der Waals and electrostatic forces. 
The fluctuation-dissipation theorem for the fluctuations of the displacements $u_i^{\rm fl}$ inside each body reads~\cite{volokitin19,volokitin20}
\begin{equation}\begin{split}
\label{eq:fluctuationdissipation_ph}
	\langle u_i^{\rm fl}(\mathbf k,\omega)[u_j^{\rm fl}(\mathbf k',&\omega')]^*\rangle=\hbar\;\mathrm{Im} M_i (k,\omega)\coth\left(\frac{\hbar \omega}{2 k_{\rm B}T_i}\right)\\
	&\,\times (2\pi)^3\delta(\mathbf k-\mathbf k')\delta(\omega-\omega')\delta_{ij},
\end{split}\end{equation}
where $\delta(x)$ is the Dirac delta distribution, $\delta_{ij}$ is the Kronecker delta, $M_i(k,\omega)$ are the mechanical susceptibilities for each body defined in Eq.~\eqref{eq:mechanicalM}, and $(\mathbf k,\omega)$-variables are related to the real space displacements by a Fourier transform,
\begin{equation}
	u_i(x,y,t)=2\,\mathrm{Re}\int_0^{\infty}\frac{\mathrm d \omega}{2\pi}\int\frac{\mathrm d^2\mathbf{k}}{(2\pi)^2}u_i(\mathbf k, \omega) \mathrm{e}^{\mathrm i (\mathbf k \cdot \mathbf r-\omega t)}.
\end{equation}
The mechanical forces per unit surface $f_i$ acting on surfaces 1 and 2 are obtained from derivatives of the interacting energy, integrating atom-wise over the two surfaces~\cite{pendry16,pendry17,volokitin19,volokitin20}, given in terms of the displacements by 
\begin{subequations}
\label{eq:f1_f2}
\begin{equation}
	f_1(\mathbf k,\omega)=au_1(\mathbf k,\omega)-bu_2(\mathbf k,\omega),
\end{equation}
\begin{equation}
	f_2(\mathbf k,\omega)=au_2(\mathbf k,\omega)-bu_1(\mathbf k,\omega),
\end{equation}
\end{subequations}
where $a$ and $b$ are defined in Eq.~\eqref{eq:a_and_b}.\\

The mechanical susceptibilities $M_i(\omega,k)$ relate the induced component of the displacements to the acting forces, such that ($i=1,2$)
\begin{equation}\label{eq:fluctuating_u}
	u_i(\mathbf k,\omega)=u_i^{\mathrm{fl}}(\mathbf k, \omega)+M_i(k,\omega)f_i(\mathbf k,\omega).
\end{equation}
In order to obtain the heat flux carried by the phonons, we need to calculate the mean power emitted by unit area given by 
\begin{equation}
\label{eq:flux_ph_app}\begin{split}
	\Phi^{(\rm ph)}&=\langle \dot{u}_2(x,y,t) f_2(x,y,t)\rangle-\langle \dot{u}_1(x,y,t) f_1(x,y,t)\rangle\\
	&=2 \int_0^\infty \frac{\mathrm{d}\omega}{2\pi}\int \frac{\mathrm{d}^2\mathbf{k}}{(2\pi)^2}\int_0^\infty \frac{\mathrm{d}\omega'}{2\pi}\int \frac{\mathrm{d}^2\mathbf{k}'}{(2\pi)^2} \omega\\
	&\,\times \mathrm{Im}[\langle u_2(\mathbf k,\omega)f^*_2(\mathbf k',\omega')\rangle -\langle u_1(\mathbf k,\omega)f^*_1(\mathbf k',\omega')\rangle].
\end{split}\end{equation}
Solving eqs.~\eqref{eq:f1_f2} and \eqref{eq:fluctuating_u}, for $f_i(\mathbf k,\omega)$ and $u_i(\mathbf k,\omega)$ in terms of $u_i^{\rm fl}(\mathbf k,\omega)$, allows one to use the fluctuation-dissipation theorem \eqref{eq:fluctuationdissipation_ph} together with \eqref{eq:flux_ph_app} to recover the Eq.~\eqref{eq:flux_ph} of the phonon flux found in the main text.

		
\section{Comparison with numerical results of Refs.~\cite{Francoeur1} and \cite{guo22}}	\label{app:comparison}
\begin{figure}[thb]
	\includegraphics[width=\linewidth]{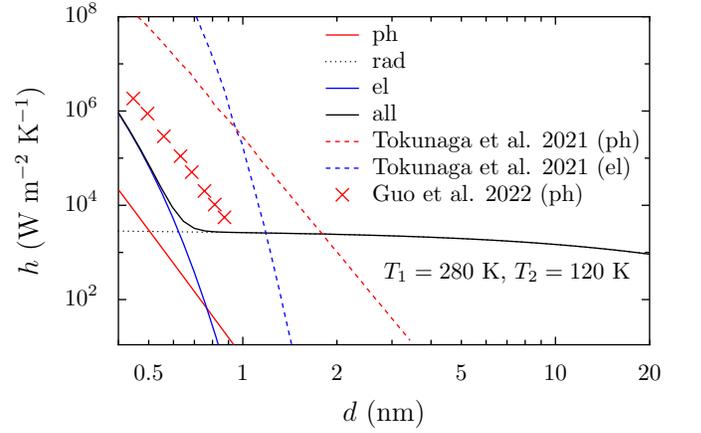}
	\caption{\label{fig:comparison}%
		Phononic and electronic thermal conductances $h=\Phi_1/(T_1-T_2)$ between two semi-infinite gold slabs as a function of the separation distance at $V_{\rm b}=0$. The results for phonons within the fluctuation-acoustodynamics approach based on continuous elastic theory (red solid line) are compared to the atomistic approaches of Refs.~\cite{Francoeur1} (red dashed lines) and Ref.~\cite{guo22} (red crosses) at 300 K. The results we obtained for electrons (blue solid line) are compared to the ones obtained within the WKB approximation in Ref.~\cite{Francoeur1}, using a different definition of the electronic heat flux. For reference, the dotted black line corresponds to the local radiative thermal conductance.}
\end{figure}

In Fig.~\ref{fig:comparison} we compare the phononic and electronic heat fluxes for $T_1=280\,K$ and $T_2=120\,$K and zero applied bias obtained within our formalism to the ones calculated in Refs.~\cite{Francoeur1} (phonons and electrons) and \cite{guo22} (phonons only, results at 300\,K) using atomistic calculations. The comparison shows that our results can be various orders of magnitude smaller. Concerning phonons, it has been argued~\cite{guo22} that this discrepancy is due to the fact that the continuous model only takes into account the forces between the elements of the surface and neglects the elements inside the material, included in the atomistic model. However, as the molecular simulations consider only a small number of atoms, it remains unclear if this prediction hold for systems with larger dimensions. As for electrons, the discrepancy is certainly due to the different definition of the heat flux, but could also be due to the simplified classical image potential used in Ref.~\cite{Francoeur1}, along with the WKB approximation.
		
\section{Thomas--Fermi electronic barrier}
\label{app:TFapp}

In this appendix, we provide the explicit expression of the regularized potential in Eq.~\eqref{eq:regularizedU} and provide a few steps for its derivation coming from Refs.~\cite{ilchenko80,ilchenko01}.

In order to obtain the regularized electronic barrier for the plane--plane configuration, it is necessary to solve the one-dimensional nonlocal Poisson's equation in the presence of a polarizable media, i.e.
\begin{equation}\begin{split}
	\left(\frac{\partial^2 }{\partial z^2}-k^2\right) G(k;z,z') - &\int\mathrm{d} z^{\prime \prime} \Pi(k;,z,z')G(q; z^{\prime\prime},z')\\
	&=\delta(z-z'),
\end{split}\end{equation}
where $G(k;z,z')$ is the Green function and $\Pi(k;z,z')$ is the polarization operator. 
In the case of three regions, allowing for specular reflection and continuity of the potential, we can write the polarization operator for a three layer system as
\begin{subequations}
\begin{equation}
	\Pi(k;z,z')=\begin{cases}
	\Pi_1(k;z-z')+\Pi_1(k;z+z'),\\
	\hspace{4cm} z,z'\leq 0,\\
	\Pi_2(k;z-z')+\Pi_2(k;z+z'),\\
	\hspace{4cm} z,z'\geq d,\\
	\Pi_{\rm gap}(k;z-z')+\Pi_{\rm gap}(k;z+z'),\\
	\hspace{4cm} 0 < z,z' <d,
	\end{cases}
\end{equation} 
since
\begin{equation}
	\Pi_s(k;z\mp z')=\int_{-\infty}^\infty \frac{\mathrm{d}q_z}{2\pi} K^2[\epsilon_s(K)-1]\exp(\mathrm i q_z[z\mp z']),
\end{equation}
\end{subequations}
where $s=1,2$ corresponds to the regions inside media 1 and 2, and $s=\mathrm{gap}$ is the region between the two media.
Here we take the static approximation $\epsilon_s(K)=\epsilon_{\rm TF}(K)$ for $s=1,2$ in Eq.~\eqref{eq:epsilon_TF} and $\epsilon_{\rm gap}=1$.
In the case of classical metals of vanishing screening length, i.e. $k_{\rm TF}\to \infty$, we recover the classical image potential of Eq.~\eqref{eq:image_potential}.\\

By calculating the diagonal self-interacting element $G(k;z)=G(k;z,z)$, we can recover the electronic barrier $U(z)$ in the absence of bias voltage corresponding to $V_{\rm b}=0$ for the three different regions as
\begin{equation}
	U_s(z)=\frac{e^2}{4\pi \epsilon_0} \left\{\frac{k_{\rm TF}}{2}- \int_0^\infty \mathrm{d}k\,k\left[G_s(k;z)+\frac{1}{2k}\right]\right\},
\end{equation}
where we add the factor $e^2k_{\rm TF}/8\pi \epsilon_0$ to set the zero energy reference inside the metal.
In the case where there is no bias, the equations are symmetric with respect to the center of the gap $z=d/2$, and
\begin{subequations} 
\begin{align}
	G_1(k;z)=&\frac{a^2_0(k,z)}{B_0(k)}[a_{\rm S}(k,0)+a_{\rm A}(k,0)+2a_0(k, 0)]\\
	\nonumber&-\frac{a_0(k,0)+a_0(k,2z)}2,
\end{align}
\begin{equation}
	G_{2}(k;z)=G_1(k;d/2-z),
\end{equation}
\begin{equation}\begin{split}
	G_{\rm gap}(k;z)&=\frac{1}{2}\left[\frac{a^2_{\rm S}(k,z)}{a_{\rm S}(k,0)+a_0(k,0)}+\frac{a^2_{\rm A}(k,z)}{a_{\rm A}(k,0)+a_0(k,0)}\right]\\
	&\,-\frac{a_{\rm A}(k, 0) + a_{\rm A}(k, 2z) + a_{\rm S}(k, 0) + a_{\rm S}(k, 2z)}{4},
\end{split}\end{equation}
\end{subequations}
where we have defined
\begin{subequations}
\begin{equation}
	B_0(k)=2[a_0(k, 0) + a_{\rm A}(k, 0)][a_0(k, 0) + a_{\rm S}(k, 0)],
\end{equation}
\begin{equation}
	a_0(k,z)=\frac{\exp(-\sqrt{k^2+k^2_{\rm TF}}|z|)}{\sqrt{k^2+k^2_{\rm TF}}},
\end{equation}

\begin{equation}
	a_{\rm S}(k,z)=\frac{\cosh(kd/2 - k\,\mathrm{mod}(z,d))}{k \sinh(kd/2)},
\end{equation}
and
\begin{equation}
	a_{\rm A}(k,z)=\operatorname{sign}\bigl(d-\mathrm{mod}(z,2d)\bigr)\frac{\sinh(kd/2 - k\,\mathrm{mod}(z,d))}{k \cosh(kd/2)},
\end{equation}
\end{subequations}
where $\operatorname{sign}(z)$ is the sign of $z$ and $\mathrm{mod}(z,d)$ is the operation $z$ modulo $d$. Taking $k_{\rm TF}\to \infty$ one recovers the expressions for the classical image potential in Eq.~\eqref{eq:image_potential}.\\

For small biases compared to the chemical potential, we are allowed to add the contribution of $V_{\rm b}$ linearly as in the expressions given in Eq.~\eqref{eq:regularizedU} in the main text.



\begin{thebibliography}{}


\bibitem{Polder}
D. Polder and M. Van Hove, \textit{Theory of Radiative Heat Transfer between Closely Spaced Bodies},
\href{http://dx.doi.org/10.1103/PhysRevB.4.3303}{Phys. Rev. B \textbf{4}, 3303 (1971)}.
%
\bibitem{Joulain_rev} K. Joulain, J.-P. Mulet, F. Marquier, R. Carminati, and J.-J. Greffet, \textit{Surface electromagnetic waves thermally excited: Radiative heat transfer, coherence properties and Casimir forces revisited in the near field}, \href{https://www.sciencedirect.com/science/article/pii/S0167572905000105}{Surf. Sci. Rep. \textbf{57}, 59 (2005)}.
%
\bibitem{Volokitin_rev} A. I. Volokitin and B. N. J. Persson, \textit{Near-field radiative heat transfer and noncontact friction}, \href{https://journals.aps.org/rmp/abstract/10.1103/RevModPhys.79.1291}{Rev. Mod. Phys. \textbf{79}, 1291 (2007)}.
%
\bibitem{Biehs_prl} S.-A. Biehs, M. Tschikin, and P. Ben-Abdallah, \textit{Hyperbolic Metamaterials as an Analog of a Blackbody in the Near Field}, \href{https://journals.aps.org/rmp/abstract/10.1103/RevModPhys.79.1291}{Phys. Rev. Lett. \textbf{109}, 104301 (2012)}.
%
\bibitem{RMP} S.-A. Biehs, R. Messina, P. S. Venkataram, A. W. Rodriguez, J. C. Cuevas, and P. Ben-Abdallah, \textit{Near-field radiative heat transfer in many-body systems}, \href{https://journals.aps.org/rmp/abstract/10.1103/RevModPhys.93.025009}{Rev. Mod. Phys., {\bf 93}, 025009 (2021)}.

\bibitem{Hu} L. Hu, A. Narayanaswamy, X. Chen, and G. Chen, \textit{Near-field thermal radiation between two closely spaced glass plates exceeding Planck’s blackbody radiation law}, \href{https://aip.scitation.org/doi/10.1063/1.2905286}{Appl. Phys. Lett. \textbf{92}, 133106 (2008)}.
%
\bibitem{Shen} S. Shen, A. Narayanaswamy, and G. Chen, \textit{Surface phonon polaritons mediated energy transfer between nanoscale gaps}, \href{https://pubs.acs.org/doi/10.1021/nl901208v}{Nano Lett. \textbf{9}, 2909–2913 (2009)}.
%
\bibitem{Rousseau} E. Rousseau, A. Siria, G. Jourdan, S. Volz, F. Comin, J. Chevrier, and J.-J. Greffet, \textit{Radiative heat transfer at the nanoscale}, \href{https://www.nature.com/articles/nphoton.2009.144}{Nat. Photon. \textbf{3}, 514 (2009)}.
%
\bibitem{Ottens} R. S. Ottens, V. Quetschke, Stacy Wise, A. A. Alemi, R. Lundock, G. Mueller, D. H. Reitze, D. B. Tanner, and B. F. Whiting, \textit{Near-field radiative heat transfer between macroscopic planar surfaces}, \href{https://journals.aps.org/prl/abstract/10.1103/PhysRevLett.107.014301}{Phys. Rev. Lett. \textbf{107}, 014301 (2011)}.
%
\bibitem{Kralik} T. Kralik, P. Hanzelka, M. Zobac, V. Musilova, T. Fort, and M. Horak, \textit{Strong near-field enhancement of radiative heat transfer between metallic surfaces}, \href{https://journals.aps.org/prl/abstract/10.1103/PhysRevLett.109.224302}{Phys. Rev. Lett. \textbf{109}, 224302 (2012)}.


\bibitem{Latella} I. Latella, S.-A. Biehs, and P. Ben-Abdallah, \textit{Smart thermal management with near-field thermal radiation}, \href{https://opg.optica.org/oe/fulltext.cfm?uri=oe-29-16-24816&id=453448}{Opt. Express \textbf{29}, 16, 24816-24833 (2021)}.
\bibitem{Fan1} K. Chen, P. Santhanam, S. Sandhu, L. Zhu, and S. Fan, \textit{Heat-flux control and solid-state cooling by regulating chemical potential of photons in near-field electromagnetic heat transfer}, \href{https://journals.aps.org/prb/abstract/10.1103/PhysRevB.91.134301}{Phys. Rev. B \textbf{91}, 134301 (2015)}.
%
\bibitem{Reddy1} L. Zhu, A. Fiorino, D. Thompson, R. Mittapally, E. Meyhofer, and P. Reddy , \textit{Near-field photonic cooling through control of the chemical potential of photons}, \href{https://www.nature.com/articles/s41586-019-0918-8}{Nature \textbf{566}, 239 (2019)}.
%
\bibitem{Srituravanich} W. Srituravanich, N. Fang, C. Sun, Q. Luo, and X. Zhang, \textit{Plasmonic Nanolithography}, \href{https://www.doi.org/10.1021/nl049573q}{Nano Lett. \textbf{4}, 1085 (2004)}.
%
\bibitem{pba_prl} P. Ben-Abdallah, \textit{Multitip Near-Field Scanning Thermal Microscopy}, \href{https://journals.aps.org/prl/abstract/10.1103/PhysRevLett.123.264301}{Phys. Rev. Lett. \textbf{123}, 264301 (2019)}.
\bibitem{De Wilde} Y. De Wilde, F. Formanek, R. Carminati, B. Gralak, P.-A. Lemoine, K. Joulain, J.-P. Mulet, Y. Chen, and J.-J. Greffet, \textit{Thermal radiation scanning tunnelling microscopy}, \href{https://www.nature.com/articles/nature05265}{Nature \textbf{444}, 740 (2006)}.
%
\bibitem{Jones} A. C. Jones, and M. B. Raschke, \textit{Thermal Infrared Near-Field Spectroscopy}, \href{https://pubs.acs.org/doi/10.1021/nl204201g}{Nano Lett. \textbf{12}, 1475 (2012)}.
\bibitem{DiMatteo} R. S. DiMatteo, P. Greiff, S. L. Finberg, K. A. Young-Waithe, H. K. H. Choy, M. M. Masaki, and C. G. Fonstad, \textit{Enhanced photogeneration of carriers in a semiconductor via coupling across a nonisothermal nanoscale vacuum gap}, \href{https://aip.scitation.org/doi/10.1063/1.1400762}{Appl. Phys. Lett. \textbf{79}, 1894 (2001)}.
%
\bibitem{Narayanaswamy} A. Narayanaswamy and G. Chen, \textit{Surface modes for near field thermophotovoltaics}, \href{https://aip.scitation.org/doi/10.1063/1.1575936}{Appl. Phys. Lett. \textbf{82}, 3544 (2003)}.
\bibitem{Laroche} M. Laroche, R. Carminati, and J.-J. Greffet, \textit{Near-field thermophotovoltaic energy conversion}, \href{https://aip.scitation.org/doi/10.1063/1.2234560}{Appl. Phys. \textbf{100}, 063704 (2006)}.
%
\bibitem{Park} K. Park, S. Basu, W. P. King, and Z. M. Zhang, \textit{Performance analysis of near-field thermophotovoltaic devices considering absorption distribution}, \href{https://www.sciencedirect.com/science/article/pii/S0022407307002397}{J. Quant. Spectros. Radiat. Transfer \textbf{109}, 305 (2008)}.
%
\bibitem{Latella2} I. Latella and P. Ben-Abdallah, \textit{Graphene-based autonomous pyroelectric system for near-field energy conversion}, \href{https://www.nature.com/articles/s41598-021-98656-8}{Sci. Rep. \textbf{11}, 19489 (2021)}.


\bibitem{kittel_2017} K. Kloppstech, N. K\"onne, S.-A. Biehs, A. W. Rodriguez, L. Worbes, D. Hellmann, and A. Kittel, \textit{Giant heat transfer in the crossover regime between conduction and radiation}, \href{http://dx.doi.org/10.1038/ncomms14475}{Nat. Commun. \textbf{8}, 14475 (2017)}.
%
\bibitem{reddy_17} L. Cui, J. Womho, V. Fernández-Hurtado, J. Feist, F. J. García-Vidal, J. C. Cuevas, E. Meyhofer, and P. Reddy, \textit{Study of radiative heat transfer in Ångström- and nanometre-sized gaps}, \href{http://dx.doi.org/10.1038/ncomms14479}{Nat. Commun. \textbf{8}, 14479 (2017)}.

\bibitem{Messina_arxiv} R. Messina, S.-A. Biehs, T. Ziehm, A. Kittel, and P. Ben-Abdallah, \textit{Heat transfer between two metals through subnanometric vacuum gaps}, \href{https://arxiv.org/abs/1810.02628}{arXiv:1810.02628 (2018)}.
%
\bibitem{Francoeur1} T. Tokunaga, A. Jarzembski, T. Shiga, K. Park, and M. Francoeur, \textit{Extreme near-field heat transfer between gold surfaces}, \href{https://journals.aps.org/prb/abstract/10.1103/PhysRevB.104.125404}{Phys. Rev. B \textbf{104}, 125404 (2021)}.
%
\bibitem{Francoeur2} T. Tokunaga, M. Arai, K. Kobayashi, W. Hayami, S. Suehara, T. Shiga, K. Park, and M. Francoeur, \textit{First-principles calculations of phonon transport across a vacuum gap}, \href{https://journals.aps.org/prb/abstract/10.1103/PhysRevB.105.045410}{Phys. Rev. B
\textbf{105}, 045410 (2022)}.
%
\bibitem{guo22} Y. Guo, C. Adessi, M.Cobian, and S. Merabia, \textit{Atomistic simulation of phonon heat transport across metallic vacuum nanogaps}, \href{https://journals.aps.org/prb/abstract/10.1103/PhysRevB.106.085403}{Phys. Rev. B \textbf{106} 085403 (2022)}.

%
\bibitem{xu}
J. B. Xu, K. L\"{a}uger, R. M\"{o}ller, K. Dransfeld, and I. H. Wilson, \textit{Energy-exchange processes by tunneling electrons}, \href{https://doi.org/10.1007/BF00332209}{App. Phys. A \textbf{7}, 155 (1994)}.

\bibitem{Sonnichsen2001} C. Sönnichsen, \textit{Plasmons in metal nanostructures}, PhD thesis University of Munich, 2001.
%
\bibitem{johnson1972} P. B. Johnson and R. W. Christy, \textit{Optical constants of the noble metals}, \href{http://dx.doi.org/10.1103/PhysRevB.6.4370}{Phys. Rev. B \textbf{6}, 4370 (1972)}.
%
\bibitem{ashcroft}N. W. Ashcroft and N. D. Mermin, \textit{Solid State Physics}, (Harcourt, Orlando, FL 1976).


\bibitem{joulain_springmodel}
Y. Ezzahri and K. Joulain, \textit{Vacuum-induced phonon transfer between two solid dielectric materials: Illustrating the case of Casimir force coupling}, \href{http://dx.doi.org/10.1103/PhysRevB.90.115433}{Phys. Rev. B \textbf{90}, 115433 (2014)}.
%
\bibitem{pendry16}
J. B. Pendry, K. Sasihithlu, and R. V. Craster, \textit{Phonon-assisted heat transfer between vacuum-separated surfaces}, \href{http://dx.doi.org/10.1103/PhysRevB.94.075414}{Phys. Rev. B \textbf{94}, 075414 (2016)}.
%
\bibitem{pendry17}
K. Sasihithlu, J. B. Pendry, and R. V. Craster, \textit{Van der Waals Force Assisted Heat Transfer}, \href{http://dx.doi.org/10.1515/zna-2016-0361}{Z. Naturforsch. \textbf{72}, 181 (2017)}.
%
\bibitem{volokitin19}
A. I. Volokitin, \textit{Effect of Resonant Photon Emission in Radiative Heat Transfer and Generation}, \href{http://dx.doi.org/10.1134/S0021364019180103}{JETP Lett. \textbf{110}, 397–404 (2019)}.
%
\bibitem{volokitin20}
A. I. Volokitin, \textit{Effect of an Electric Field in the Heat Transfer between Metals in the Extreme Near Field}, \href{https://doi.org/10.1088/1361-648X/ab71a5}{J. Phys.: Condens. Matter \textbf{32} 215001 (2020)}.
%
\bibitem{persson}
B. N. J. Persson, \textit{Theory of rubber friction and contact mechanics}, \href{https://doi.org/10.1063/1.1388626}{J. Chem. Phys. \textbf{115} 3840 (2001)}.
%
\bibitem{hamaker}
P. Pinchuk and K. Jiang, \textit{Size-dependent Hamaker Constants for Silver and Gold Nanoparticles}, \href{https://doi.org/10.1117/12.2187282}{Proc. of SPIE \textbf{9549}, 95491 (2015)}.
%

\bibitem{fordweber}
G. W Ford and W. H. Weber, \textit{Electromagnetic interactions of molecules with metal surfaces}, \href{http://dx.doi.org/10.1016/0370-1573(84)90098-X}{Phys. Rep. \textbf{113}, 195 (1984)}.
%
\bibitem{kittel_2005}
A. Kittel, W. Müller-Hirsch, J. Parisi, S.-A. Biehs, D. Reddig, and M. Holthaus, \textit{Near-Field Heat Transfer in a Scanning Thermal Microscope}, \href{http://dx.doi.org/10.1103/PhysRevLett.95.224301}{Phys. Rev. Lett. \textbf{95}, 224301 (2005)}.
%
\bibitem{poc}
P.-O. Chapuis, S. Volz, C. Henkel, K. Joulain, and J.-J. Greffet, \textit{Effects of spatial dispersion in near-field radiative heat transfer between two parallel metallic surfaces}, \href{http://dx.doi.org/10.1103/PhysRevB.77.035431}{Phys. Rev. B \textbf{77}, 035431 (2008)}.


\bibitem{simmons}
J. G. Simmons, \textit{Generalized Formula for the Electric Tunnel Effect between Similar Electrodes Separated by a Thin Insulating Film}, \href{https://doi.org/10.1063/1.1702682}{J. Appl. Phys \textbf{34}, 1793 (1963)}.
%
\bibitem{WKB}
M. V. Berry and K. E. Mount, \textit{Semiclassical approximations in wave mechanics}, \href{https://doi.org/10.1088/0034-4885/35/1/306}{Rep. Prog. Phys. \textbf{35}, 315 (1972)}.
%
\bibitem{impenetrable}
M. Andrews, \textit{Singular potentials in one dimension}, \href{https://doi.org/10.1119/1.10585}{Am. J. Phys. \textbf{44}, 1064 (1976)}.
%
\bibitem{bardeen}
J. Bardeen, \textit{Theory of the Work Function. II. The Surface Double Layer}, \href{https://doi.org/10.1103/PhysRev.49.653}{Phys. Rev. \textbf{49}, 653 (1936)}.
%
\bibitem{eguiluzPRL}
A.G. Eguiluz, M. Heinrichsmeier, A. Fleszar, and W. Hanke, \textit{First-Principles Evaluation of the Surface Barrier for a Kohn-Sham Electron at a Metal Surface}, \href{https://doi.org/10.1103/PhysRevLett.68.1359}{Phys. Rev. Lett. \textbf{68}, 1359 (1992)}.
%
\bibitem{Tmatrixmethod}
Y. Ando and I. Tomohiro, \textit{Calculation of transmission tunneling current across arbitrary potential barriers}, \href{https://doi.org/10.1063/1.338082}{J. App. Phys. \textbf{61}, 1497 (1987)}.

%
\bibitem{hishinuma}
Y. Hishinuma, T. H. Geballe, B. Y. Moyzhes, and T. W. Kenny, \textit{Refrigeration by combined tunneling and thermionic emission in vacuum: Use of nanometer scale design}, \href{https://doi.org/10.1063/1.1365944}{App. Phys. Lett. \textbf{78}, 2572 (2001)}.
%
\bibitem{ilchenko80}
A.M. Gabovich, L. G. Il'chenko, E. A. Pashitskii, and Yu. A. Romanov, \textit{Electrostatic energy and screened charge interaction near the surface of metals with different Fermi surface shape}, \href{https://doi.org/10.1016/0039-6028(80)90163-6}{Surf. Sci. \textbf{13}, 179 (1980)}.
%
\bibitem{ilchenko01}
L.G. Il'chenko and T.V. Goraychuk, \textit{Role of the image forces potential in the formation of the potential barrier between closely spaced metals}, \href{https://doi.org/10.1016/S0039-6028(01)00844-5}{Surf. Sci. \textbf{478}, 169-179 (2001)}.
%
\bibitem{imry82}
U. Sivan and Y. Imry, \textit{Multichannel Landauer formula for thermoelectric transport with application to thermopower near the mobility edge}, \href{https://doi.org/10.1103/PhysRevB.33.551}{Phys. Rev. B \textbf{33}, 551 (1986)}.
%
\bibitem{datta92} R. Lake and S. Datta, \textit{Energy balance and heat exchange in mesoscopic systems}, \href{https://doi.org/10.1103/PhysRevB.46.4757}{Phys. Rev. B \textbf{46}, 4757 (1992)}.
%
\bibitem{buttiker11} R. S\'{a}nchez and M. B\"{u}ttiker, \textit{Optimal energy quanta to current conversion}, \href{https://doi.org/10.1103/PhysRevB.83.085428}{Phys. Rev. B \textbf{83}, 085428 (2011)}.
\end{thebibliography}
\end{document}